\pdfoutput=1
\documentclass{article}
\usepackage[a4paper]{geometry}
\usepackage{graphicx}
\usepackage{amssymb,amsmath,amsthm}
\usepackage{tikz}
\usepackage{subcaption}
\usepackage{authblk}
\usepackage{appendix}
\usepackage[labelfont={bf,small},textfont={small}]{caption}
\usepackage{natbib}
\usepackage{hyperref}

\theoremstyle{definition}

\renewcommand{\vec}[1]{{\boldsymbol{#1}}}
\renewcommand{\epsilon}{\varepsilon}

\title{Seismic metasurfaces: Sub-wavelength resonators and Rayleigh wave interaction}
\author[1]{D. J. Colquitt\footnote{Corresponding author: \href{mailto:d.colquitt@liverpool.ac.uk}{d.colquitt@liverpool.ac.uk}}}
\author[2]{A. Colombi}
\author[2]{R. V. Craster}
\author[3]{P. Roux}
\author[4]{S. R. L. Guenneau}
\affil[1]{Department of Mathematical Sciences, University of Liverpool, Liverpool, L69 7ZL, UK}
\affil[2]{Department of Mathematics, Imperial College London, London, SW7 2AZ, UK}
\affil[3]{ISTerre, Observatoire de Grenoble, Universit\'e de Grenoble 1 Joseph-Fourier, Grenoble, France}
\affil[4]{Aix-Marseille Univ., CNRS, Centrale Marseille, Institut Fresnel, UMR 7249, 13397 Marseille Cedex 20, France}

\usepackage{tikz}
\usetikzlibrary{decorations.pathmorphing}
\usetikzlibrary{decorations.markings}
\usepackage{sidecap}
\usepackage{subcaption}
\usepackage{authblk}
\usepackage{booktabs}

\makeatletter\newcommand*{\at}{@}\makeatother

\input epsf 
\usepackage{amsmath}

  {\left\lbrace\begin{array}{@{}l@{}}}%
  {\end{array}\right.}
  
\newcommand{\be}{\begin{equation}}
\newcommand{\ee}{\end{equation}}
\newcommand{\beq}{\begin{equation}}
\newcommand{\eeq}{\end{equation}}

\long\def\symbolfootnote[#1]#2{\begingroup%
\def\thefootnote{\fnsymbol{footnote}}\footnote[#1]{#2}\endgroup}

\renewcommand{\vec}[1]{{\boldsymbol{#1}}}
\renewcommand{\epsilon}{\varepsilon}
\newcommand{\dd}{\mathrm{d}}
\newcommand{\pd}{\partial}
\newcommand{\uu}[1]{\underline{\underline{#1}}}
\newcommand{\Rho}{\mathrm{P}}
\DeclareMathOperator{\sech}{sech}

\begin{document}

\date{\today}

\maketitle

\label{firstpage}

\begin{abstract}
We consider the canonical problem of an array of rods, which act as resonators, placed on an elastic substrate; the substrate being either a thin elastic plate or an elastic half-space.
In both cases the flexural plate, or Rayleigh surface, waves in the substrate interact with the resonators to create interesting effects such as effective band-gaps for surface waves or filters that transform surface waves into bulk waves; these effects have parallels in the field of optics where such sub-wavelength resonators create metamaterials, and metasurfaces, in the bulk and at the surface respectively.

Here we carefully analyse this canonical problem by extracting the dispersion relations analytically thereby examining the influence of both the flexural and compressional resonances on the propagating wave.
For an array of resonators atop an elastic half-space we augment the analysis with numerical simulations.
Amongst other effects, we demonstrate the striking effect of a dispersion curve that transitions from Rayleigh wave-like to shear wave-like behaviour and the resultant change in displacement from surface to bulk waves.  
\end{abstract}

\section{Introduction}

Metamaterials, as synthetic composite
materials with a structure such that they exhibit properties not
usually found in natural materials, now form a major emerging research
area that barely existed before 2000; in fact, the term
``Metamaterial'' itself was first used in 1999. Since then, the area
has grown extensively and shows little sign of slowing down.  The key
point is that materials can be designed to have, say, a negative
refractive index as predicted by~\cite{veselago1968electrodynamics} and later by\cite{pendry00a}; and subsequently fabricated by
\cite{smith00a}, which is impossible in naturally occurring
materials. The first metamaterials were developed in optics and
electromagnetism and relied upon having simultaneously negative
permittivity and permeability, this was made physically possible using
a microstructured periodic medium consisting of sub-wavelength resonators such
as split-ring resonators \cite[]{lagarkov1997resonance,pendry99a}. It has subsequently been
realised that these ideas can also be profitably utilised to create acoustic or
elastic (negative density and negative shear or bulk modulus)
metamaterials \cite[]{craster12a,deymier13a,liu2000locally,fang2006ultrasonic,kadic2013metamaterials}. Similarly,
although metamaterials were initially developed for bulk media, one
can also create microstructured surfaces that act as metasurfaces 
\cite[]{maradudin11a} with, again, most of the activity centered around
electromagnetic waves and surface plasmons.
Given that the array of
resonators we introduce here modifies the surface wave properties we
choose to call this a metasurface rather than a metamaterial although
there is no strict definition, as yet.

Much more recently seismic metamaterials have begun
to be considered, although here the challenges are substantial: not
only are the Rayleigh waves, which are of primary interest, surface
waves, but the underlying system of equations are the full vector
equations of elasticity. Nonetheless there have been attempts to
modify the local properties of the ground through the addition of
inclusions, or resonators, of a different material at a sub-wavelength
scale. The different types of inclusions, resonant or non-resonant,
determine the properties and the performance of the structured medium or
metamaterial/ metasurface.  Turning to the non-resonant case first, \cite{brule14a}
show with both large-scale experiments and theory that the soil
properties can be critically affected by periodic arrangements of
boreholes with a spacing of about a metre; in the Bragg scattering regime these induce bandgaps that
can be used for seismic protection.  More recently, resonant
sub-wavelength scatterers placed on top of an elastic substrate have been investigated and we are motivated
by recent geophysical experiments \cite[]{colombi15b} that have
demonstrated that natural forest trees can act as a metasurface for
frequencies between 30 and 100 Hz. Quite remarkably these experiments
show a distinct reduction in the transmission of waves over a broad
range of frequencies; these frequencies appear to lie in bandgaps
created by local resonances between trees and elastic waves in the
substrate.  These large-scale experiments can be interpreted
through laboratory-scale experiments and theory
\cite[]{colombi14a,williams15a,yoritomo2016band} that have, for simplicity, utilised
elastic plates, not the full elastic system; notably, these papers neglect the flexural deformations of the resonators.
In contrast, in the present paper, we account for both the flexural and compressional deformations of the resonators and demonstrate that the flexural resonances significantly impact the spectrum of the metamaterial in certain regimes.
Although we have focused
upon the geophysical setting, elastic waves are also very important in
ultrasonics and surface acoustic wave devices, and the Bragg scattering
aspects of surface arrays as phononic crystals have been explored,
i.e., in, say, \cite{achaoui11a}.
Locally resonant structures have also been shown to give rise to so-called \emph{super-wide pseudo-directional} band gaps in platonic crystals~\cite[]{xiao2012flexural}; such band gaps are formed by the coalescence of Bragg and resonance band gaps.

Given the interest in this emerging area, and the developing
applications such as surface to bulk elastic wave filters \cite[]{colombi16a}, there is a need for the fundamental solutions to
canonical problems involving periodic arrays of sub-wavelength
resonators atop elastic substrates. Here we provide the required
analytical and theoretical background for these seismic metasurfaces
by considering arrays of these resonators attached to elastic substrates, thereby
considering the interaction with elastic Rayleigh waves and for
comparison and completeness we also consider the elastic plate. We do
so in a two-dimensional setting as the essential concepts are
uncluttered by excessive algebra, and the solutions we obtain allow
for interpretation and are, on occasion, completely explicit.

The mathematical tools for treating periodic arrays are extensive:
linear partial differential operators, quasi-periodic Green's functions, and
integral transforms based upon Fourier series and we use a variety of
these approaches freely to obtain the cleanest solutions;
these theoretical approaches are complimented by illustrative numerical calculations   based on the Spectral Element Method.
 We begin in section \ref{sec:plates} by considering the array of resonators attached to an
 elastic plate, this is complementary to \cite{williams15a}, and we
 obtain explicit dispersion relations and explore the relative
 importance of flexural and compressional resonances in the array and
 how they interact with the plate waves. This naturally leads in the
 full elastic situation of an array atop an elastic half-space,
 section \ref{sec:full_elastic} investigates this and again explicit
 results emerge that are used to interpret and predict metamaterial/metasurface
 phenomena. Finally, we draw together some concluding remarks in section \ref{sec:conclude}.

\section{Thin elastic plates}
\label{sec:plates}

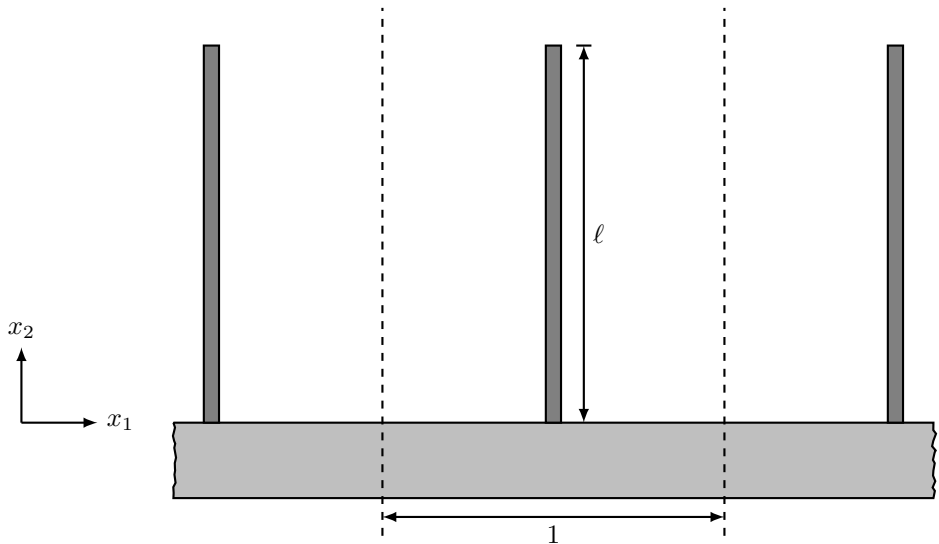
\begin{figure}
\centering
\begin{tikzpicture}
\draw[thick,-latex] (-7 , 0 ) -- (-6 , 0);
\draw[thick,-latex] (-7 , 0 ) -- (-7 , 1);
\node[right] at (-6,0) {$\displaystyle x_1$};
\node[above] at (-7,1) {$\displaystyle x_2$};
\draw[thick,fill=lightgray] (-5,0) -- (5,0) decorate[decoration={random steps,segment length=2,amplitude=1}]{ -- (5,-1)} -- (-5,-1) decorate[decoration={random steps,segment length=2,amplitude=1}]{ -- (-5,0)};
\draw[thick,fill=gray] (-0.1 , 0 ) rectangle (0.1 , 5);
\draw[thick,fill=gray] (4.4 , 0 ) rectangle (4.6 , 5);
\draw[thick,fill=gray] (-4.4 , 0 ) rectangle (-4.6 , 5);
\draw[thick,latex-latex] (0.4 , 0 ) -- (0.4 , 5);
\draw[thick] (0.3 , 5 ) -- (0.5 , 5);
\node[right] at ( 0.4 , 2.5 ) {$\displaystyle \ell$};
\draw[thick, dashed] (-2.25 , -1.5 ) -- (-2.25 , 5.5 );
\draw[thick,latex-latex] (-2.25 , -1.25 ) -- (2.25 , -1.25);
\node[below] at ( 0 , -1.25 ) {$\displaystyle 1$};
\draw[thick, dashed] (2.25 , -1.5 ) -- (2.25 , 5.5 );
\end{tikzpicture}
\caption{\label{fig:plate-schematic}
The schematic representation of the linear array of resonators
$\mathcal{R}_n$ attached to the surface of a thin elastic plate $\mathcal{G}_n$.
The boundaries of the elementary cell are indicated by the vertical dashed lines.}
\end{figure}

We begin by considering an infinite periodic array of beams (which support both flexural and compressional waves, whereas rods only support compressional waves) on a thin elastic plate, as shown in
figure~\ref{fig:plate-schematic}; the boundary of the elementary
cell is indicated by the dashed lines. We use Cartesian coordinates
$x_1, x_2$ orientated along the thin elastic plate, and perpendicular
to it with $u_1, u_2$
being the associated components of the displacement vector. 
It is convenient to introduce two sets corresponding to the two
different components within each cell. 
The first defines the flexural foundation $\mathcal{G}_n = \{\vec{x} : |x_1| < (n+1)/2 , x_2 = 0\}$, whilst the second corresponds to the resonator $\mathcal{R}_n = \{\vec{x} : x_1 = n , 0 < x_2 < \ell \}$ for $n\in\mathbb{Z}$.
We note that $\mathcal{G}_n\cap\mathcal{R}_m = \emptyset$ and $\bar{\mathcal{G}}_n\cap\bar{\mathcal{R}}_n = \{\vec{x} : \vec{x} = (n,0) \}$.

The equations of motion for time-harmonic waves of radian frequency $\omega$ are
\begin{subequations}
\label{eq:plate-eom}
\begin{alignat}{4}
\left(\frac{\pd^4}{\pd x_1^4} - \beta^4\omega^2\right)u_2(x_1) & =
\frac{1}{EI}\sum_{n\in\mathbb{Z}} \left(
V_n\delta(x_1-n) + M_m\delta^\prime(x_1 - n) \right)
\quad\text{for}\quad\vec{x}\in\mathcal{G}_n,\\
\left(\frac{\pd^2}{\pd x_1^2} + \alpha^2\omega^2\right)u_1(x_1) & =
\frac{1}{E}\sum_{n\in\mathbb{Z}}
F_n\delta(x_1-n)
\quad\text{for}\quad\vec{x}\in\mathcal{G}_n,\\
\left(\frac{\pd^4}{\pd x_2^4} - \beta_R^4\omega^2\right)u_1(x_2) & = 0
\quad\text{for}\quad\vec{x}\in\mathcal{R}_n,\\
\left(\frac{\pd^2}{\pd x_2^2} + \alpha_R^2\omega^2\right)u_2(x_2) & = 0
\quad\text{for}\quad\vec{x}\in\mathcal{R}_n,
\end{alignat}
\end{subequations}
where $\beta^4 = \rho h/(EI)$, $\alpha^2 = \rho/E$, $E$ is the
Young's modulus of the foundation, $I$ is the area moment of inertia
of the foundation, and $h$ is the thickness of the foundation; 
 the subscript $R$ denotes the corresponding properties for the resonators.
The amplitudes of the horizontal and vertical forces and moments at
the base of the resonators are denoted by $F_n$, $V_n$, and $M_n$
respectively; the Dirac delta function, $\delta$, and its derivative,
$\delta^\prime$, are understood in the distributional sense.

The tops of the resonators are free leading to boundary conditions of the form
\begin{equation}
\frac{\pd^2 u_1}{\pd x_2^2} = \frac{\pd^3 u_1}{\pd x_2^3} = \frac{\pd u_2}{\pd x_2} = 0 \quad\text{for}\quad\vec{x}\in\mathcal{R}_n\cup\{\vec{x} : x_2 = \ell\}.
\end{equation}
Imposing continuity of forces and moments at the base of the resonators yields
\begin{equation}
\frac{\pd^2 u_1}{\partial x_2^2} = -\frac{M_n}{EE_RI_R},\quad
\frac{\pd^3 u_1}{\partial x_2^3} = -\frac{F_n}{EE_RI_R},\quad
\frac{\pd u_2}{\partial x_2} = -\frac{V_n}{ESE_RI_R},
\label{eq:plate-interface-cond}
\end{equation}
for $\vec{x}\in\mathcal{R}_n\cup\{\vec{x} : x_2 = 0\}$ where $S$ is the cross-sectional area of the resonators.
These boundary conditions are then supplemented with kinematic equations corresponding to continuity of displacements and rotations at the base of the resonator
\begin{equation}
u_1(n,0^-) = - u_1(n,0^+),\;
u_2(n,0^-) = u_2(n,0^+),\;
\frac{\pd}{\pd x_1} u_2(n,0^-) = -\frac{\pd}{\pd x_1} u_1(n,0^+),
\end{equation}
where $x_2 = 0^-$ belongs to the closure of $\mathcal{G}_n$ and $x_2 = 0^+$ belongs to the closure of $\mathcal{R}_n$.
The negative signs in the first and final continuity conditions are a result of the choice of coordinate system and forcing orientation.

Introducing the Fourier transform and its inverse as
\begin{equation}
u^\mathrm{F}(k) = \int\limits_{-\infty}^\infty u(x_1) e^{-ikx_1} \dd x\quad,\quad
u(x_1) = \frac{1}{2\pi}\int\limits_{-\infty}^\infty u^\mathrm{F}(k) e^{ikx_1} \dd k,
\end{equation}
respectively, 
together with the Bloch-Floquet quasi-periodicity conditions on the forces and moments $(\cdot)_n  = (\cdot)e^{i\xi n}$ where $\xi\in(-\pi,\pi)$, the fields are expressed in the form
\begin{subequations}
\begin{equation}
u_1(x_1) = \frac{F}{2\pi}\sum_{n\in\mathbb{Z}}\;\int\limits_{-\infty}^\infty
\frac{e^{in(\xi-k)}e^{ikx_1}}{\alpha^2\omega^2 - k^2} \dd k,
\end{equation}
\begin{equation}
u_2(x_1) = \frac{1}{2\pi}\sum_{n\in\mathbb{Z}}\; \int\limits_{-\infty}^\infty
\frac{[V - ikM]e^{in(\xi-k)}e^{ikx_1}}{k^4 - \beta^4\omega^2} \dd k,
\end{equation}
\end{subequations}
where we have restricted ourselves to $x_2 = 0$ and the second argument is therefore omitted.
Making use of the Poisson summation formula~\citep{lighthill58a},
\begin{equation}
\sum_{n\in\mathbb{Z}} f(2\pi ) = \frac{1}{2\pi}\sum_{m\in\mathbb{Z}}\; \int\limits_{-\infty}^\infty f(n) e^{imn} \dd n,
\end{equation}
the fields can be further reduced to infinite sums.
Setting $x_1 = 0$ leads to a $3\times3$ linear system for the displacements and rotations, the solvability condition for which yields the dispersion equation
\begin{equation}
\label{eq:plate-disp}
\det\left( \uu{\sigma} - \uu{I} \right) = 0,
\end{equation}
where $\uu{I}$ is the identity matrix and
\begin{equation}
\label{eq:plate-sigma}
\uu{\sigma} = \sum_{n\in\mathbb{Z}}
\begin{bmatrix}
V/D^{(p)}_n &
-iM_\theta(\xi - 2\pi n)/D^{(p)}_n &
-iM_u(\xi - 2\pi n)/D^{(p)}_n \\
-iV(\xi - 2\pi n)/D^{(p)}_n &
M_\theta(\xi - 2\pi n)^2/D^{(p)}_n &
M_u(\xi - 2\pi n)^2/D^{(p)}_n \\
0 &
F_\theta/D^{(m)}_n &
F_u/D^{(m)}_n
\end{bmatrix},
\end{equation}
with $D^{(m)}_n = \alpha^2\omega^2 - (\xi - 2\pi n)^2$ and $D^{(p)}_n = (\xi - 2\pi n)^4 - \beta^4\omega^2$ being the dispersion equations for a uniform infinite membrane and plate respectively,
\begin{subequations}
\label{eq:plate-forces-moments}
\begin{equation}
\label{eq:plate-force-v}
V = \omega \lambda_R E_R S_R\tan(\ell\lambda_R\omega)/(EI),
\end{equation}
\begin{equation}
\label{eq:plate-moment-theta}
M_\theta = \frac{(1+i)\beta_RE_RI_R\sqrt{\omega}}{2EI}\frac{\sin\left[ (1+i)\beta_R\ell\sqrt{\omega}\right] - \sinh\left[ (1+i)\beta_R\ell\sqrt{\omega}\right]}{1+\cos\left(\beta_R\ell\sqrt{\omega}\right)\cosh\left(\beta_R\ell\sqrt{\omega}\right)},
\end{equation}
\begin{equation}
\label{eq:plate-moment-u}
M_u = \frac{\beta^2E_RI_R\omega}{EI}\frac{\sin\left(\beta_R\ell\sqrt{\omega}\right)\sinh\left(\beta_R\ell\sqrt{\omega}\right)}{1+\cos\left(\beta_R\ell\sqrt{\omega}\right)\cosh\left(\beta_R\ell\sqrt{\omega}\right)},
\end{equation}
\begin{equation}
\label{eq:plate-force-theta}
F_\theta = -\frac{\beta^2E_RI_R\omega}{E}\frac{\sin\left(\beta_R\ell\sqrt{\omega}\right)\sinh\left(\beta_R\ell\sqrt{\omega}\right)}{1+\cos\left(\beta_R\ell\sqrt{\omega}\right)\cosh\left(\beta_R\ell\sqrt{\omega}\right)},
\end{equation}
and
\begin{equation}
\label{eq:plate-force-u}
F_u = -\frac{(1-i)\beta_R^3E_RI_R\omega^{3/2}}{2E}\frac{\sin\left[ (1+i)\beta_R\ell\sqrt{\omega}\right] - \sinh\left[ (1+i)\beta_R\ell\sqrt{\omega}\right]}{1+\cos\left(\beta_R\ell\sqrt{\omega}\right)\cosh\left(\beta_R\ell\sqrt{\omega}\right)}.
\end{equation}
\end{subequations}
It now remains to evaluate the infinite sums appearing in the matrix problem.

\subsection{Evaluation of the infinite sums}
We start by considering the simplest summation which, using Poisson summation, can be expressed as
\begin{equation}
\sum_{n\in\mathbb{Z}} \frac{1}{D^{(p)}_n} =
\frac{1}{2\pi}\sum_{n\in\mathbb{Z}}e^{in\xi}\int\limits_{-\infty}^\infty\frac{e^{in\gamma}}{\gamma^4 - \beta^4\omega^2}\dd\gamma.
\end{equation}
The integrand has four simple poles located at $\gamma = \pm\beta\sqrt{\omega}$ and $\gamma = \pm i\beta\sqrt{\omega}$.
In order to evaluate the integral we take the usual semi-circular contour closing it in the upper half-plane for $n\geq0$ and the lower half-plane otherwise.
For $n\geq0$ the contour is indented such that the pole at $\gamma = -\beta\sqrt{\omega}$ lies outside the contour and $\gamma = \beta\sqrt{\omega}$ is enclosed by the contour.
The converse is true for $n < 0$.
Computing the residues we find that
\begin{equation}
\sum_{n\in\mathbb{Z}} \frac{1}{D^{(p)}_n} = \frac{1}{4\beta^3\omega^{3/2}}\left(
\frac{\sinh\beta\sqrt{\omega}}{\cos\xi - \cosh\beta\sqrt{\omega}} - \frac{\sin\beta\sqrt{\omega}}{\cos\xi - \cos\beta\sqrt{\omega}}\right).
\end{equation}
For the remaining summations we find that
\begin{equation}
\sum_{n\in\mathbb{Z}} \frac{(\xi-2\pi n)}{D^{(p)}_n} = \frac{i\sin\xi(\cosh\beta\sqrt{\omega} - \cos\beta\sqrt{\omega})}{4\beta^2\omega(\cos\xi - \cosh\beta\sqrt{\omega})(\cos\xi - \cosh\beta\sqrt{\omega})},
\end{equation}
\begin{equation}
\sum_{n\in\mathbb{Z}} \frac{(\xi-2\pi n)^2}{D^{(p)}_n} = \frac{1}{4\beta\sqrt{\omega}}\left(
\frac{\sinh\beta\sqrt{\omega}}{\cos\xi - \cosh\beta\sqrt{\omega}} + \frac{\sin\beta\sqrt{\omega}}{\cos\xi - \cos\beta\sqrt{\omega}}\right),
\end{equation}
and finally
\begin{equation}
\sum_{n\in\mathbb{Z}} \frac{1}{D^{(m)}_n} = \frac{\sin\alpha\omega}{2\alpha\omega(\cos\xi - \cos\alpha\omega)}.
\end{equation}

\subsection{The dispersion equation}

\begin{figure}
\centering
\begin{subfigure}[t]{0.49\linewidth}
\includegraphics[width=\linewidth]{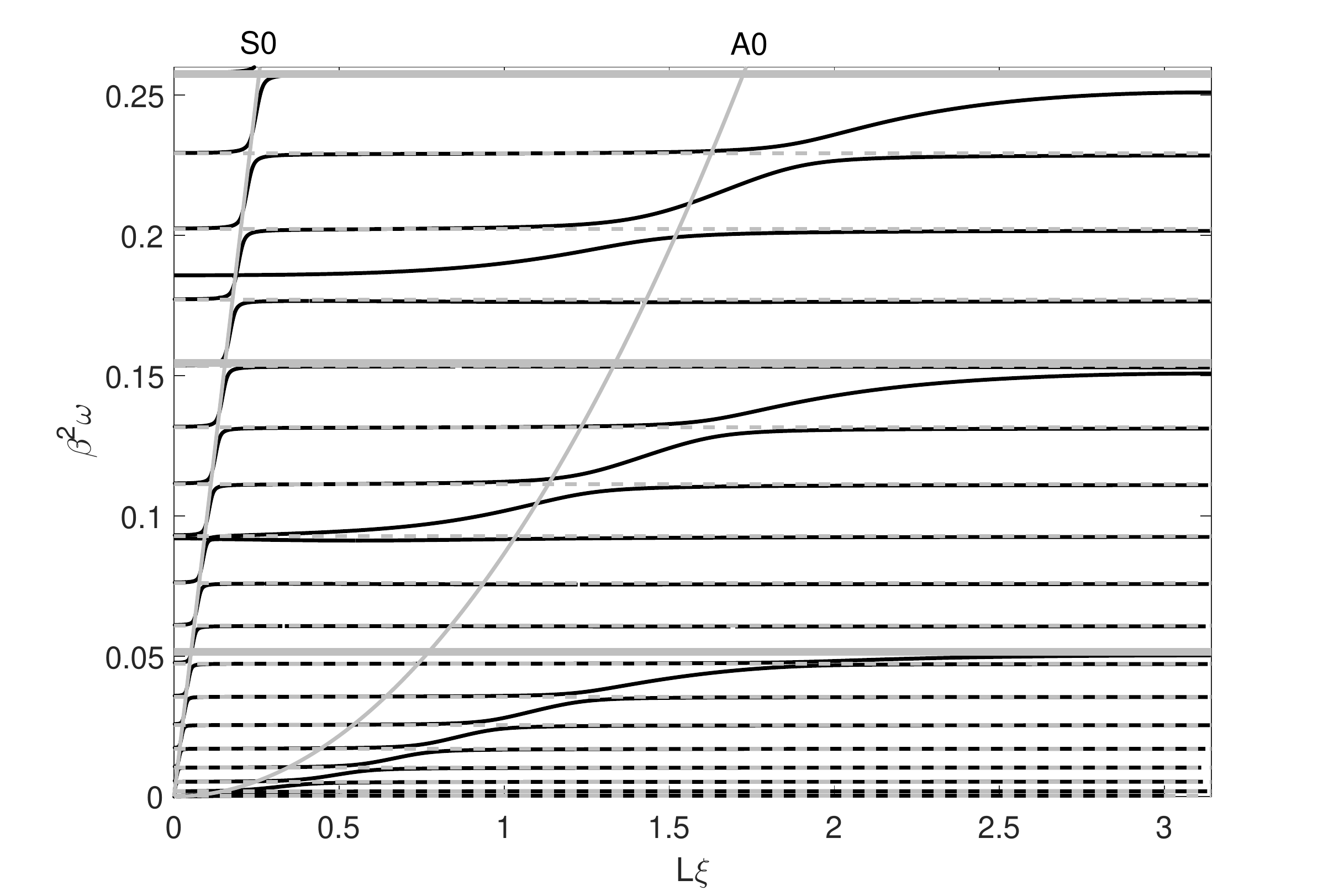}
\caption{\label{fig:plate-full-disp}}
\end{subfigure}
\begin{subfigure}[t]{0.49\linewidth}
\includegraphics[width=\linewidth]{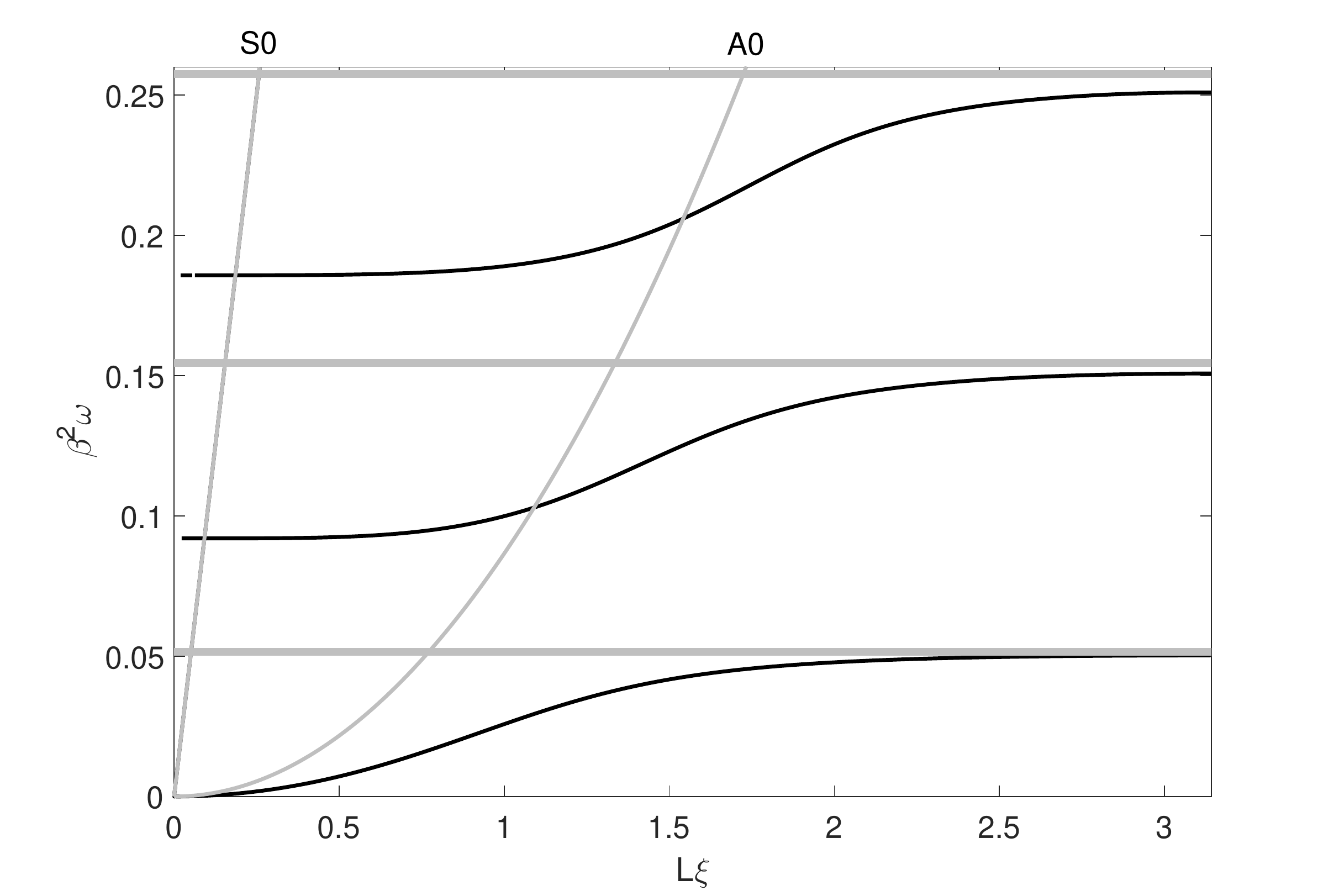}
\caption{\label{fig:plate-compressional-disp}}
\end{subfigure}
\caption{\label{fig:plate-disp}
The dispersion curves for the plate system.
Panel (a) shows the case when the flexural interactions of the resonators are accounted for; panel (b) shows the curves when we neglect these flexural interactions and consider the compressional resonator modes only.
The solid black lines show the solutions of the dispersion equation,
the thin solid grey lines indicate the compressional and flexural dispersion curves for the $S_0$ and $A_0$ modes of an elastic plate without resonators.
The dashed grey lines indicate the flexural resonances of the resonators, whilst the thick solid grey lines denote the compressional resonances of the resonators.}
\end{figure}

Now that the $3\times 3$ matrix $\uu{\sigma}$ is expressed in a finite number of terms, it is straightforward to compute the determinant~\eqref{eq:plate-disp} and hence obtain the dispersion equation
\begin{equation}
(\cos\xi - \cosh\beta\sqrt{\omega})(\cos\xi - \cos\beta\sqrt{\omega})
\left(\sum_{n=0}^3 A_n(\omega)\cos^n\xi\right) = 0.
\label{eq:plate-disp2}
\end{equation}
The coefficients $A_n(\omega)$ are given in \ref{app:dispersion-plate} and contain the information about the material and geometrical properties of the resonators.
The first parenthesised term in equation~\eqref{eq:plate-disp2} has no positive real roots for $\omega$ and therefore can be ignored; the second term is the dispersion equation for flexural waves in a uniform thin plate.

The remaining parenthesised term in equation~\eqref{eq:plate-disp2} is a cubic polynomial in $\cos\xi$ and therefore has exact closed-form solutions.
An example of a typical dispersion diagram is shown in
figure~\ref{fig:plate-disp}, the corresponding material and
geometrical parameters are detailed in
table~\ref{tab:parameters}.
The dispersion curves for compressional and flexural waves in the
foundation, without resonators, are also shown in figure~\ref{fig:plate-disp}.
The grey dashed horizontal lines are associated with the flexural resonances of the beams; the corresponding boundary value problem is that of an Euler-Bernoulli beam with one end clamped at $x_2 = 0$ and the remaining end free at $x_2 = \ell$.
In this case, the resonances satisfy the transcendental equation~\cite[]{graff75a}
\begin{equation}
\cos\beta_R\sqrt{w}\ell\cosh\beta_R\sqrt{w}\ell = -1.
\end{equation}
We note that the density of the resonances reduces with increasing frequency, tending toward the expected constant value~\cite[]{graff75a}.
The compressional resonances, indicated by the thick solid grey lines in figure~\ref{fig:plate-disp}, are associated with the spectrum of longitudinal waves in a clamped-free thin elastic rod;
the natural frequencies are thus $\omega^{(r)}_n = (2n-1)\pi/(2\ell\alpha_R)$.

\begin{table}
\centering
\begin{tabular}{c@{\qquad}c@{\qquad}c@{\qquad}}
\toprule
\multicolumn{2}{c}{Parameter} 			&			  \\
Symbol		&		 Definition 		&		 Value \\
\midrule
$L$			& Lattice spacing			& $0.02\;\text{m}$ \\
$\rho$		& Plate density				& $2700\;\text{kg}\cdot\text{m}^{-3}$\\
$h$			& Plate thickness			& $6\times10^{-3}\;\text{m}$\\
$d$			& Plate width				& $2\times10^{-2}\;\text{m}$\\
$E$			& Plate Young's modulus		& $69\;\text{GPa}$\\
$\ell$		& Resonator length			& $0.61\;\text{m}$ \\
$\rho_R$	& Resonator density			& $2700\;\text{kg}\cdot\text{m}^{-3}$\\
$h$			& Resonator diameter		& $6.35\times10^{-3}\;\text{m}$\\
$E$			& Resonator Young's modulus	& $69\;\text{GPa}$\\
\bottomrule
\end{tabular}
\caption{\label{tab:parameters}
The geometrical and numerical parameters used to produce the dispersion curves for the plate system shown in figure~\ref{fig:plate-disp}.}
\end{table}

Comparing figures~\ref{fig:plate-full-disp} and~\ref{fig:plate-compressional-disp} we observe that, away from the flexural resonances (indicated by the dashed grey lines), the spectrum appears virtually unchanged by the flexural deformations of the resonators.
This effect can be attributed to the contrast in rigidities between the plate and resonator: the plate is roughly 4.5 times as rigid as the resonators, and so the flexural deformations of the resonators couple weakly to the plate.

\begin{figure}
\centering
\begin{subfigure}[t]{0.49\linewidth}
\includegraphics[width=\linewidth]{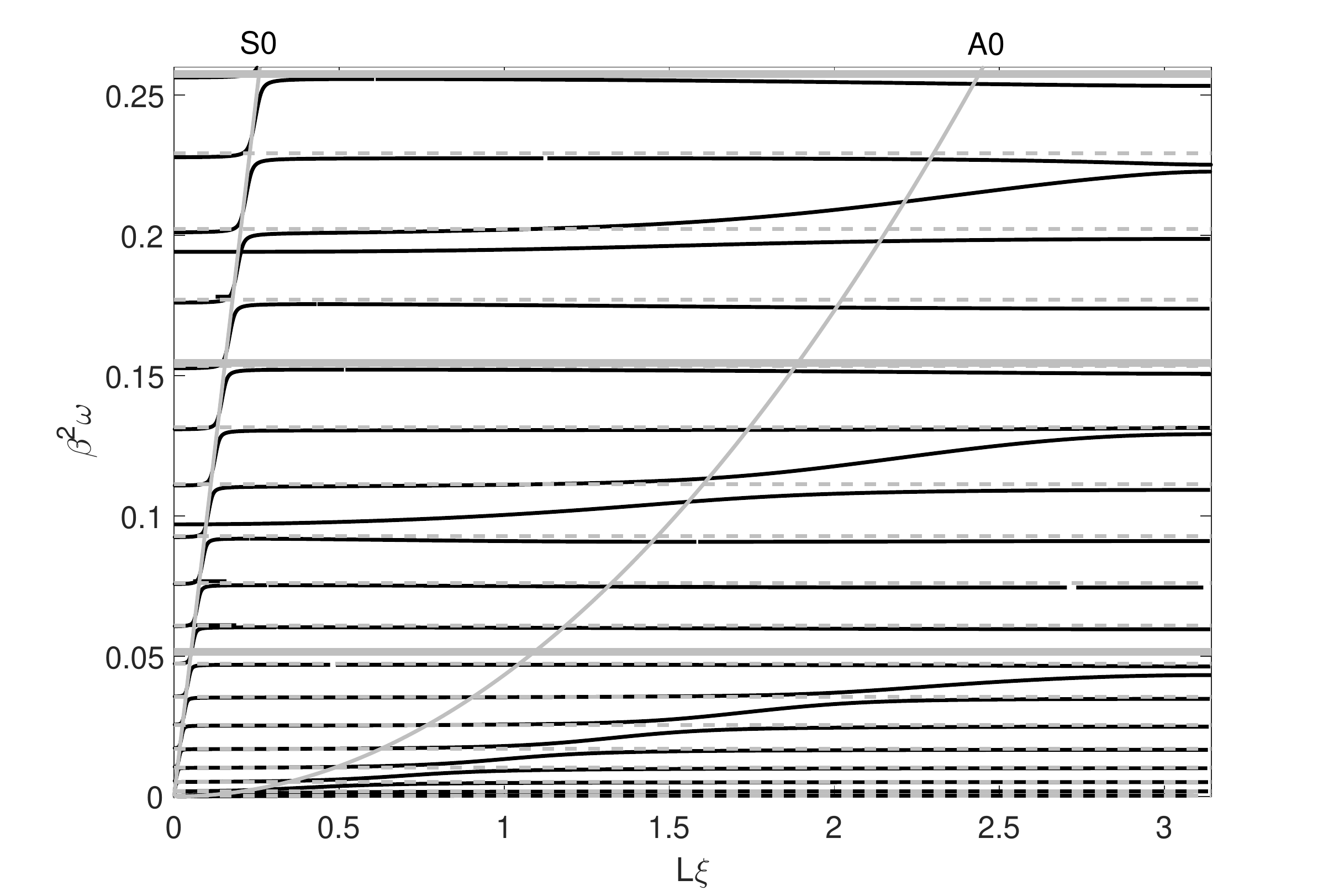}
\caption{\label{fig:plate-full-disp-thin}}
\end{subfigure}
\begin{subfigure}[t]{0.49\linewidth}
\includegraphics[width=\linewidth]{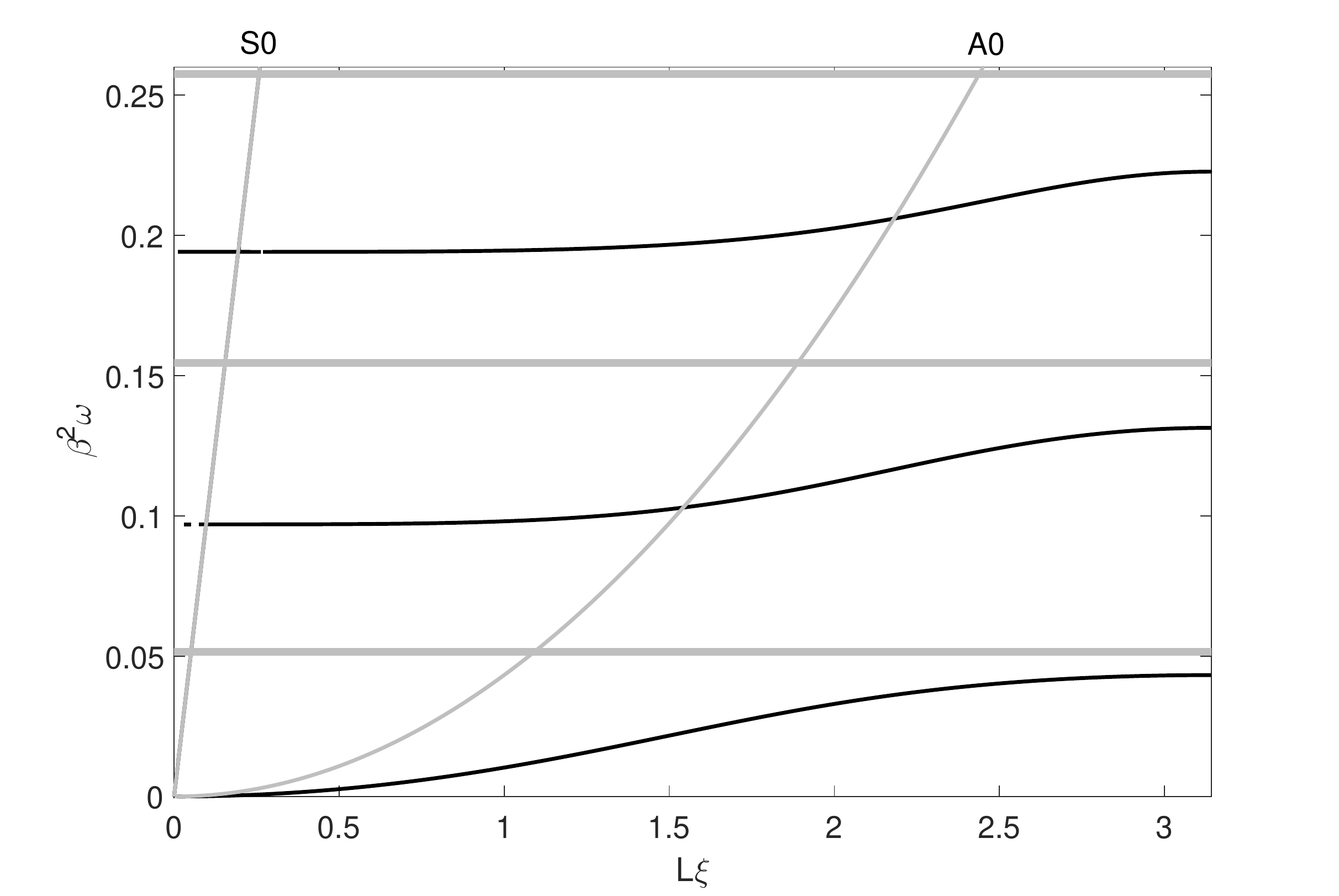}
\caption{\label{fig:plate-compressional-disp-thin}}
\end{subfigure}
\caption{\label{fig:plate-disp-thin}
The dispersion curves for the plate system with parameters identical to those for figure~\ref{fig:plate-disp} and detailed in table~\ref{tab:parameters}, except that the thickness of the plate is $3\times10^{-3}\;\text{m}$.
Panel (a) shows the case when the flexural interactions of the resonators are accounted for; panel (b) show the curves when we neglect these flexural interactions and consider the compressional resonator modes only.
The solid black lines show the solutions of the dispersion equation,
the thin solid grey lines indicate the compressional and flexural dispersion curves for the $S_0$ and $A_0$ modes of an elastic plate without resonators.
The dashed grey lines indicate the flexural resonances of the resonators, whilst the thick solid grey lines denote the compressional resonances of the resonators.}
\end{figure}

Figure~\ref{fig:plate-disp-thin} shows the dispersion curves for a system where the base plate is thinner with $h=3\times10^{-3}\;\text{m}$, but otherwise identical to that considered earlier; all the other parameters are as detailed in table~\ref{tab:parameters}.
In this configuration, the rigidity of the plate is approximately half that of the resonators and results in the flexural deformations of the resonators coupling strongly with the flexural waves in the base plate.
Indeed, comparing figures~\ref{fig:plate-disp} and~\ref{fig:plate-disp-thin}, we note that the flexural resonances (indicated by the dashed grey lines) begin to play a more important role.
For the thicker more rigid plate, the compressional resonance of the resonators give rise  to band gaps.
However, for the thinner more flexible plate, the confinement of the dispersion curves between adjacent flexural resonances gives rise to additional band gaps as well as regions of negative group velocity (c.f. the region near $\omega=0.225$ and $\xi=\pi$ in figure~\ref{fig:plate-disp-thin}); this negative group velocity can be associated with so-called \emph{Double-negative acoustic metamaterials}, as discussed by~\cite{li2004double}.
In previous works, these flexural resonances have been neglected but, as we see here they can play an important role in certain regimes.

This effect can be understood in terms of the forces and moments exerted at the junctions between the plate and resonators.
In particular, equations~\eqref{eq:plate-forces-moments} give the force (resp. moment) per unit displacement (resp. rotation) exerted by the resonators on the plate.
The compressional deformations of the resonators couple to the plate by means of the force $V$~\eqref{eq:plate-force-v}, whilst $M_\theta$~\eqref{eq:plate-moment-theta}, $M_u$~\eqref{eq:plate-moment-u}, $F_\theta$~\eqref{eq:plate-force-theta}, and  $F_u$~\eqref{eq:plate-force-u} couple the flexural deformations to the plate.
Examining equations~\eqref{eq:plate-forces-moments}, we observe that decreasing the thickness of the plate, and hence the area moment of inertia, increases the magnitude of the moments $M_\theta$ and $M_u$, which couple the flexural deformations of the resonators to the plate.

\begin{figure}
\centering
\begin{subfigure}[t]{0.49\linewidth}
\includegraphics[width=\linewidth]{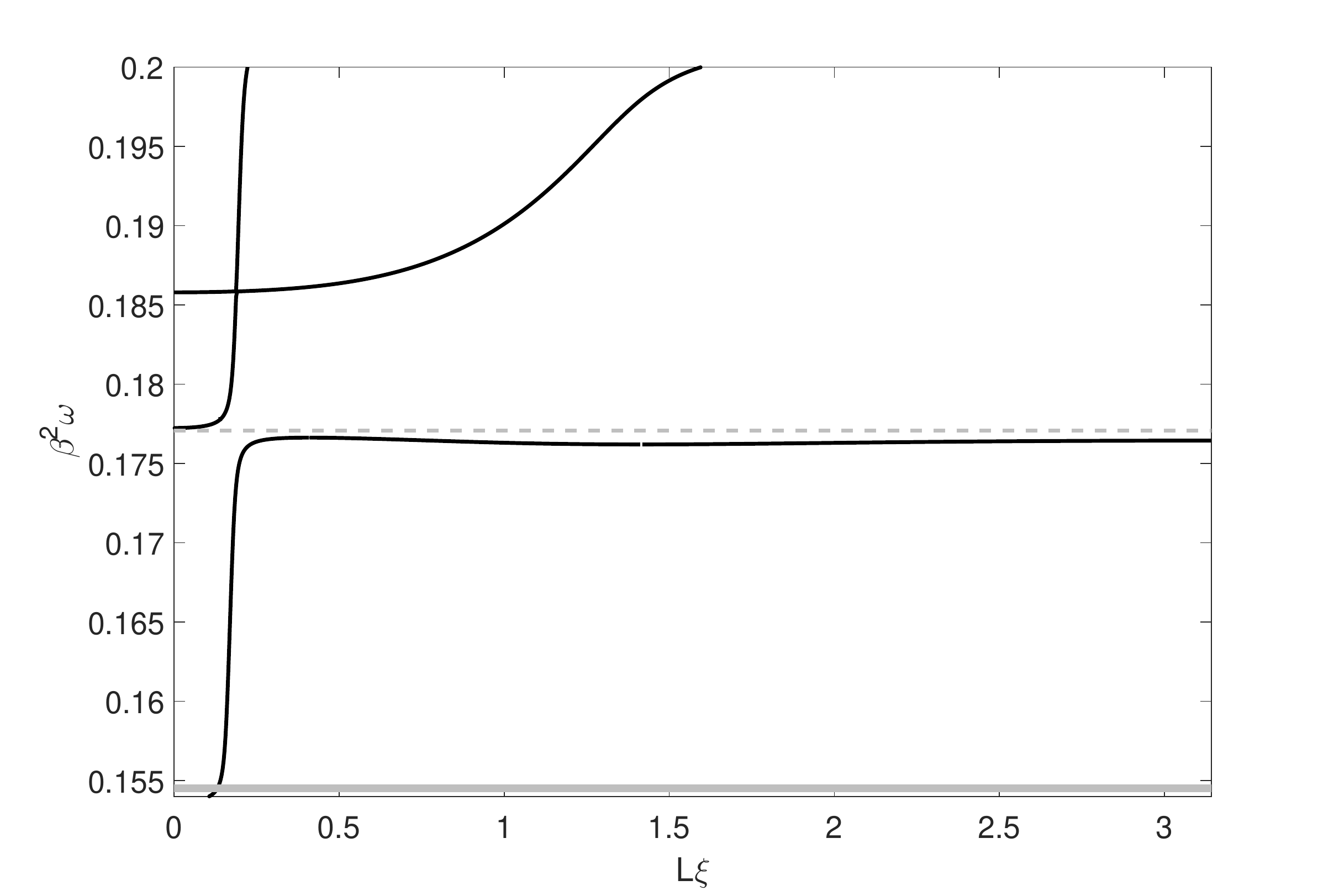}
\caption{\label{fig:plate-zoom-dispersion}}
\end{subfigure}
\begin{subfigure}[t]{0.49\linewidth}
\includegraphics[width=\linewidth]{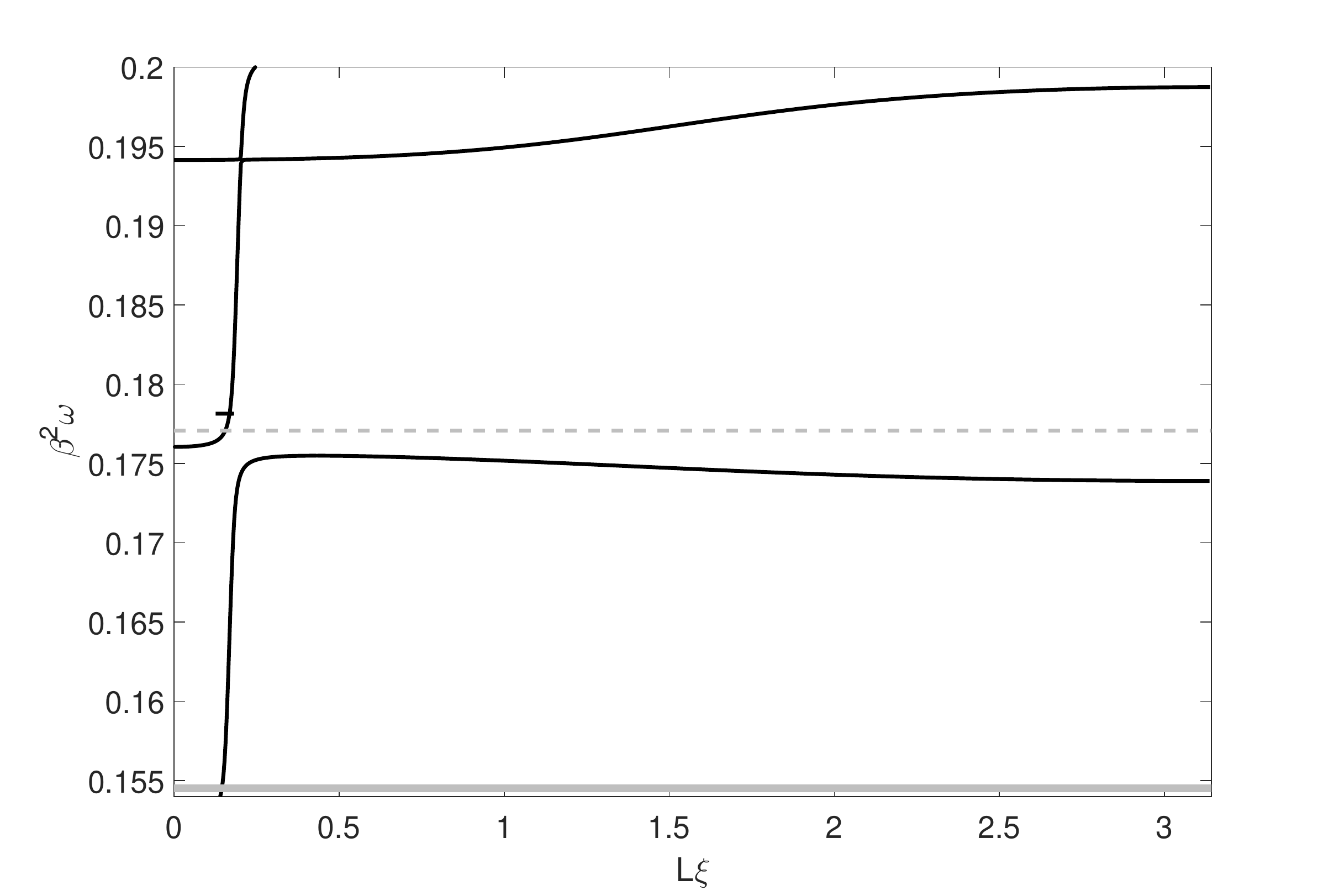}
\caption{\label{fig:plate-zoom-dispersion-thin}}
\end{subfigure}
\caption{\label{fig:plate-disp-zoom}
Magnified sections of the dispersion curves shown in figures~\ref{fig:plate-full-disp} and~\ref{fig:plate-full-disp-thin}.
Panel(a) corresponds to the dispersion curves for the thicker more rigid plate shown in figure~\ref{fig:plate-full-disp}, whilst (b) corresponds to the curves for the thinner more flexible shown ~\ref{fig:plate-full-disp-thin}.
The solid grey curve indicates the compressional resonance, whilst the dashed grey curve corresponds to the flexural resonance.}
\end{figure}

It has been observed experimentally by~\cite{roux2016new} that, for sufficiently thin plates, sharp transmission bands appear in the band gaps created by the longitudinal resonances of the resonators.
Figure~\ref{fig:plate-disp-zoom} shows a magnified region inside the first band gap of the spectra illustrated in figures~\ref{fig:plate-disp} and~\ref{fig:plate-disp-thin};
Panel (a) corresponds to the thicker more rigid plate, whilst panel (b) is associated with the thinner more flexible plate.
As can be observed from figure~\ref{fig:plate-disp-zoom}, these narrow transmission bands  (corresponding to the solid black curves) are associated with the flexural resonances of the resonators.
Specifically, the confinement of the dispersion curves between adjacent flexural resonances results in regions of the spectrum with very small, but non-zero, group velocities (the gradients of the dispersion curves); this in turn results in very narrow transmission bands inside the band gaps created by the longitudinal resonances.
As we transition from the more rigid plate, c.f. figure~\ref{fig:plate-zoom-dispersion}, to the more flexible plate, c.f. figure~\ref{fig:plate-zoom-dispersion-thin}, the group velocity and the band width increase and therefore one would expect the effect to be more noticeable in an experimental setting.
However, we emphasise that this effect persists in both the rigid and flexible case and serves to further confirm both the analysis presented here and the earlier experimental results; additionally, this effect highlights that the effects of the flexural resonances can be important, particularly for flexible plates.

\subsection{Rod-like resonators}

From the previous discussion it would appear that, for sufficiently rigid plates, the flexural component of the resonators interacts weakly with the plate and its primary influence is seen as a discrete set of resonances on the dispersion diagram.
With this in mind, we now examine the case where the flexural interaction is considered negligible.
Under this assumption we may set $F_n = M_n = 0$ in the equations of motion~\eqref{eq:plate-eom} and~\eqref{eq:plate-interface-cond} after-which the flexural and longitudinal boundary value problems decouple and we need only consider the vertical motion of the flexural foundation and resonators.
The dispersion equation is then
\begin{equation}
\label{eq:plate_disp_long}
\left(\cos\xi - \cosh\beta\sqrt{\omega}\right)
\left(\cos\xi - \cos\beta\sqrt{\omega}\right)
\left(\cos\xi - \cos\alpha\omega\right)
\left(\sum_{n=0}^2 B_n(\omega)\cos^n\xi\right) = 0,
\end{equation}
where $B_n(\omega)$ are given in ~\ref{app:dispersion-plate-long}.
As for the case of flexural resonators, the first parenthesised term has no positive real roots for $\omega$ and the second is simply the dispersion equation for flexural waves in a uniform plate.
Similarly, the third term is the dispersion equation for longitudinal waves in a thin rod; this leaves the final parenthesised term, which is quadratic in $\cos\xi$, as the dispersion equation for flexural waves travelling through the plate with resonators and solutions can again be found in closed form.
The solutions to the dispersion equation for the system of rod-like resonators are shown as the dashed red lines in figure~\ref{fig:plate-disp} where it can be seen that the flexural resonances present for the case of beam-like resonators are absent.
The remaining modes, corresponding to the compressional modes in the resonators, are present in both cases.

\section{An elastic half-plane}
\label{sec:full_elastic}

\begin{figure}
\centering
\begin{tikzpicture}
\draw[thick,-latex] (-7 , 0 ) -- (-6 , 0);
\draw[thick,-latex] (-7 , 0 ) -- (-7 , 1);
\draw[thick] (-7,0) circle (0.15);
\draw[thick,fill=black] (-7,0) circle (0.05);
\node[right] at (-6,0) {$\displaystyle x_1$};
\node[above] at (-7,1) {$\displaystyle x_2$};
\node[left,below] at (-7.15,-0.15) {$\displaystyle x_3$};
\draw[thick,fill = lightgray] (-5,0) -- (5,0) decorate[decoration={random steps,segment length=3,amplitude=2}]{arc (0:-180:5 and 2)};
\draw[thick,fill=gray] (-0.1 , 0 ) rectangle (0.1 , 5);
\draw[thick,fill=gray] (4.4 , 0 ) rectangle (4.6 , 5);
\draw[thick,fill=gray] (-4.4 , 0 ) rectangle (-4.6 , 5);
\draw[thick,latex-latex] (0.4 , 0 ) -- (0.4 , 5);
\draw[thick] (0.3 , 5 ) -- (0.5 , 5);
\node[right] at ( 0.4 , 2.5 ) {$\displaystyle \ell$};
\draw[thick, dashed] (-2.25 , -1.5 ) -- (-2.25 , 5.5 );
\draw[thick,latex-latex] (-2.25 , -1.25 ) -- (2.25 , -1.25);
\node[below] at ( 0 , -1.25 ) {$\displaystyle L$};
\draw[thick, dashed] (2.25 , -1.5 ) -- (2.25 , 5.5 );
\end{tikzpicture}
\caption{\label{fig:half-plane-schematic}
The linear array of rod-like resonators $\mathcal{R}$ on the surface of an elastic half-space $\mathcal{H}$.
The boundaries of the elementary cell, $\vert x_1\vert<L/2$, are indicated by the vertical
dashed lines.}
\end{figure}
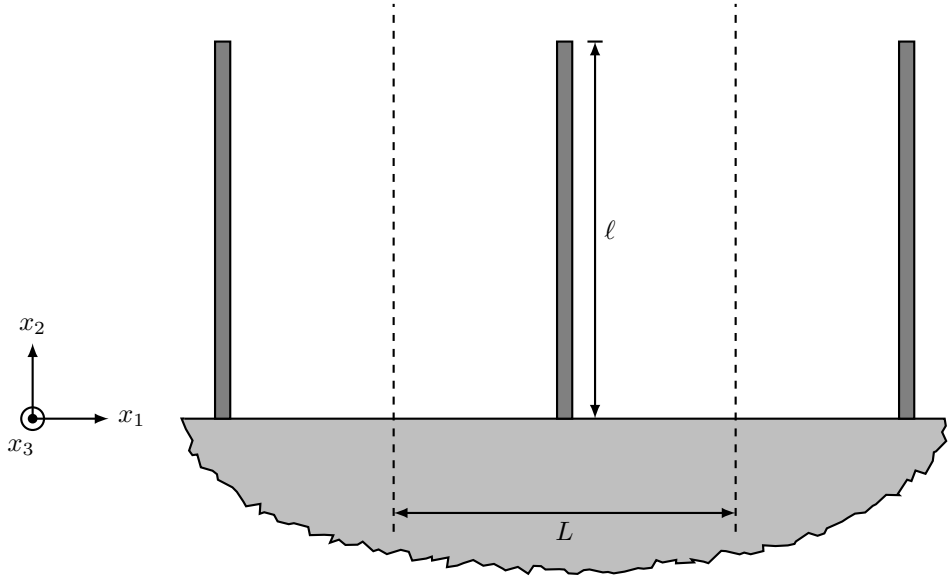

We now move to the more geophysically relevant, but involved, problem of a linear array of resonators on an elastic half-plane, as shown in figure~\ref{fig:half-plane-schematic}.
In particular, we are interested in the propagation and control of Rayleigh waves, that is, waves that propagate along the surface of the half-space and decay exponentially into the bulk.
We begin by formulating the problem as a doubly periodic two-dimensional square array of resonators on an elastic half-space and introduce the sets $\mathcal{R} = \{\vec{x}
: x_1 = 0 , 0 < x_2 < \ell , x_3 = 0 \}$ and $\mathcal{H} = \{\vec{x}
: |x_1| < L/2 , x_2 < 0 , |x_3| < L/2 \}$, which correspond to the resonator and half-space in the elementary cell.
It was shown, in \S\ref{sec:plates} as well as in previous experimental~\cite[]{colombi14a} and numerical~\cite[]{williams15a} investigations, that the contribution from the bending deformations of the resonators does not significantly affect the overall behaviour of the system away from flexural resonances.
Therefore, and in order to ease exposition, only the longitudinal
motion of the resonators will be considered; the approach used here is
equally applicable to the case when the flexural deformation of the resonators is included.
With this being the case, the boundary value problem for the displacement amplitude field $\vec{u} = \vec{u}(\vec{x})$ is expressed as
\begin{subequations}
\begin{alignat}{3}
\label{eq:elastic-eom}
\nabla\cdot\uu{C} : \nabla\vec{u} + \rho\omega^2\vec{u} & = \vec{0} \quad
\text{for}\quad \vec{x}\in\mathcal{H},\\
\label{eq:elastic-resonator-eom}
\left(\frac{\pd^2}{\pd x_2^2} + \alpha_R^2\omega^2\right)u_2(x_2) & = 0
\quad\text{for}\quad\vec{x}\in\mathcal{R},\\
\label{eq:boundary-conditions}
\left( \uu{C} : \nabla\vec{u} \right)\cdot\vec{n} -
\left[ 0 , V\delta(x_1)\delta(x_3) , 0 \right]^\mathrm{T} &  = \vec{0} \quad\text{for}\quad
x_2 = 0,
\end{alignat}
\end{subequations}
where $\uu{C}$ is the elastic stiffness tensor of the half-space and $\vec{n}$ is the outward unit normal.
The first equation~\eqref{eq:elastic-eom} is the equation of motion for bulk elastic waves in the half-space, whilst~\eqref{eq:elastic-resonator-eom} is the equation of motion of longitudinal waves in the resonators.
Finally, equation~\eqref{eq:boundary-conditions} represents the balance of forces at the junction between the resonators and half-plane; continuity of $u_2$ across $x_2=0$ at $x_1=x_3=0$ is also imposed.

For the isotropic half-space considered here, it is convenient to employ the Helmholtz decomposition and express the displacement field in terms of the usual scalar and solenoidal vector potentials \cite[]{achenbach84a}
\begin{equation}
\vec{u}(\vec{x}) = \nabla\varphi(\vec{x}) + \nabla\times\vec{\psi}(\vec{x}).
\end{equation}
The compressional and shear potentials both satisfy Helmholtz equations
\begin{subequations}
\begin{equation}
\nabla^2\phi(\vec{x}) + \Lambda_c^2\phi(\vec{x}) = 0,
\end{equation}
\begin{equation}
\nabla^2\vec{\psi}(\vec{x}) + \Lambda_s^2\vec{\psi}(\vec{x}) = \vec{0},
\end{equation}
\end{subequations}
with $\Lambda_c^2 = \rho\omega^2/(\lambda + 2\mu)$, $\Lambda_s^2 =
\rho\omega^2/\mu$, and $\lambda$ and $\mu$ being the Lam\'e parameters
of the elastic half-space.
The necessary components of the tractions, corresponding to the boundary conditions~\eqref{eq:boundary-conditions}, are then
\begin{subequations}
\begin{equation}
\sigma_{12} = \mu(2\phi_{,12} - 2\vec{\psi}_{,11} - \Lambda_s^2\vec{\psi}),
\end{equation}
\begin{equation}
\sigma_{32} = \mu(2\phi_{,32} - 2\vec{\psi}_{,33} - \Lambda_s^2\vec{\psi}),
\end{equation}
\begin{equation}
\sigma_{22} = 2\mu(\phi_{,22} - \vec{\psi}_{,12}) - \lambda\Lambda_c^2\phi.
\end{equation}
\end{subequations}
Having formulated the problem in three spatial dimensions, which was necessary in order to ensure correct dimensionality, we now focus on the problem analogous to that considered in \S\ref{sec:plates}: a linear array of resonators on a half space.
To this end, and motivated by the results of \S\ref{sec:plates}, we impose Bloch-Floquet quasi-periodicity and search for solutions in the form of Fourier series
\begin{subequations}
\begin{equation}
\phi(\vec{x}) = \sum_{n\in\mathbb{Z}} \phi_n\exp(i(k - 2 \pi n / L )x_1 + \alpha_n x_2),
\end{equation}
\begin{equation}
\vec{\psi}(\vec{x}) = \sum_{n\in\mathbb{Z}} \vec{\psi}_n\exp(i(k - 2 \pi n / L)x_1 + \beta_n x_2),
\end{equation}
\label{eq:potentials}
\end{subequations}
with $\alpha_n^2 = (k- 2\pi n/L)^2 - \Lambda_c^2$ and $\beta_n^2 = (k - 2\pi n/L)^2 - \Lambda_s^2$.
The branch cuts are chosen such that $\Re(\alpha_n) > 0$ and $\Re(\beta_n) > 0$ in order to ensure decay as $x_2\to-\infty$.
Physically, the above Ans{\"a}tze
correspond to surface waves propagating parallel to the $x_1$-axis
over the surface of an elastic half-space with a periodic array of slender rods.

Combining~\eqref{eq:boundary-conditions} and~\eqref{eq:potentials}, we find
\begin{equation}
\sum_{p\in\mathbb{Z}}\Gamma_p\phi_p e^{i(k-2\pi p / L) x_1} = -\frac{V(\omega)}{2\mu}\sum_{n\in\mathbb{Z}}\delta(x_1)\delta(x_3)\gamma_n\phi_n e^{i(k -2\pi n / L) x_1},
\label{eq:dispersion-sum}
\end{equation}
where
\[
\Gamma_p = 2(k - \pi p/L)^2 - \Lambda_s^2 -
\frac{ 4(k - 2\pi p/L)^2\sqrt{ (k - 2\pi p/L)^2 - \Lambda_c^2 }\sqrt{ (k - 2\pi p/L)^2 - \Lambda_s^2 } }{ 2(k - 2\pi p/2)^2 - \Lambda_s^2 },
\]
and
\[
\gamma_n = \Lambda_s^2\frac{ \sqrt{ (k - 2\pi n/L)^2 - \Lambda_c^2 } }{ 2(k - 2\pi n/L)^2 - \Lambda_s^2 }.
\]
Multiplying both sides of~\eqref{eq:dispersion-sum} by $e^{-i(k-2\pi m/L)x_1}$ ($m\in\mathbb{Z}$) and integrating over $(x_1,x_3)\in(-L/2,L/2)^2$ yields
\begin{equation}
\Gamma_m\phi_m = - \frac{V(\omega)}{\mu L^2}\sum_{n\in\mathbb{Z}}\gamma_n\phi_n,
\label{eq:elastic-compat}
\end{equation}
which corresponds to an infinite matrix problem with solvability condition
\begin{equation}
\det\left(
\uu{I} + \frac{V(\omega)}{\mu L^2}\uu{M}\right) = 0,
\label{eq:det-disp}
\end{equation}
with $[\uu{M}]_{mn} = \gamma_n/\Gamma_m$ and $\underline{\underline{I}}$ being the identity matrix.

Equation~\eqref{eq:det-disp} is the dispersion equation for surface waves propagating along the boundary of the half-plane.
The infinite matrices $\uu{M}$ and $\uu{I}$ can be truncated at some appropriate order and solved to give the solutions $(\omega,k)$ of the dispersion equation.
However for our purposes, i.e. the analysis of sub-wavelength control of surface waves, it is sufficient to consider a single mode expansion such that $\phi_m = \phi_0\delta_{m0}$ and $\psi_m = \psi_0\delta_{m0}$.
It is clear that solutions of this form satisfy~\eqref{eq:elastic-compat} and, hence,~\eqref{eq:det-disp}.
In this case, the dispersion equation reduces to
\begin{equation}
4\xi^2\sqrt{\xi^2 - r^2}\sqrt{\xi^2-1} - ( 2\xi^2 - 1 )^2 = \sqrt{\xi^2 - r^2}\frac{V(\omega)}{\omega L^2 \sqrt{\mu\rho}},
\label{eq:reduced-dispersion}
\end{equation}
where we recognise the left-hand side as the usual Rayleigh
 dispersion equation \cite[]{achenbach84a}, and 
where $V(\omega) = S\omega\sqrt{E\Rho}\tan(\ell\omega\sqrt{\Rho/E})$ is the vertical force exerted on the half-plane by the resonators.
We have also introduced the normalised variables $\xi = k/\Lambda_s$ and $r^2 = \Lambda_c^2/\Lambda_s^2 = 1/(2 + \lambda/\mu)$, and $0 < r^2 < 3/4$.
Although the reduced dispersion equation is transcendental in both $\xi$ and $\omega$ and does not, in general, permit closed form solutions, it can be expressed as a sixth order polynomial in $\xi^2$ allowing the roots to be determined efficiently using the various fast algorithms available for finding polynomial roots.

\begin{figure}[h]
\centering
\includegraphics[width=0.7\linewidth]{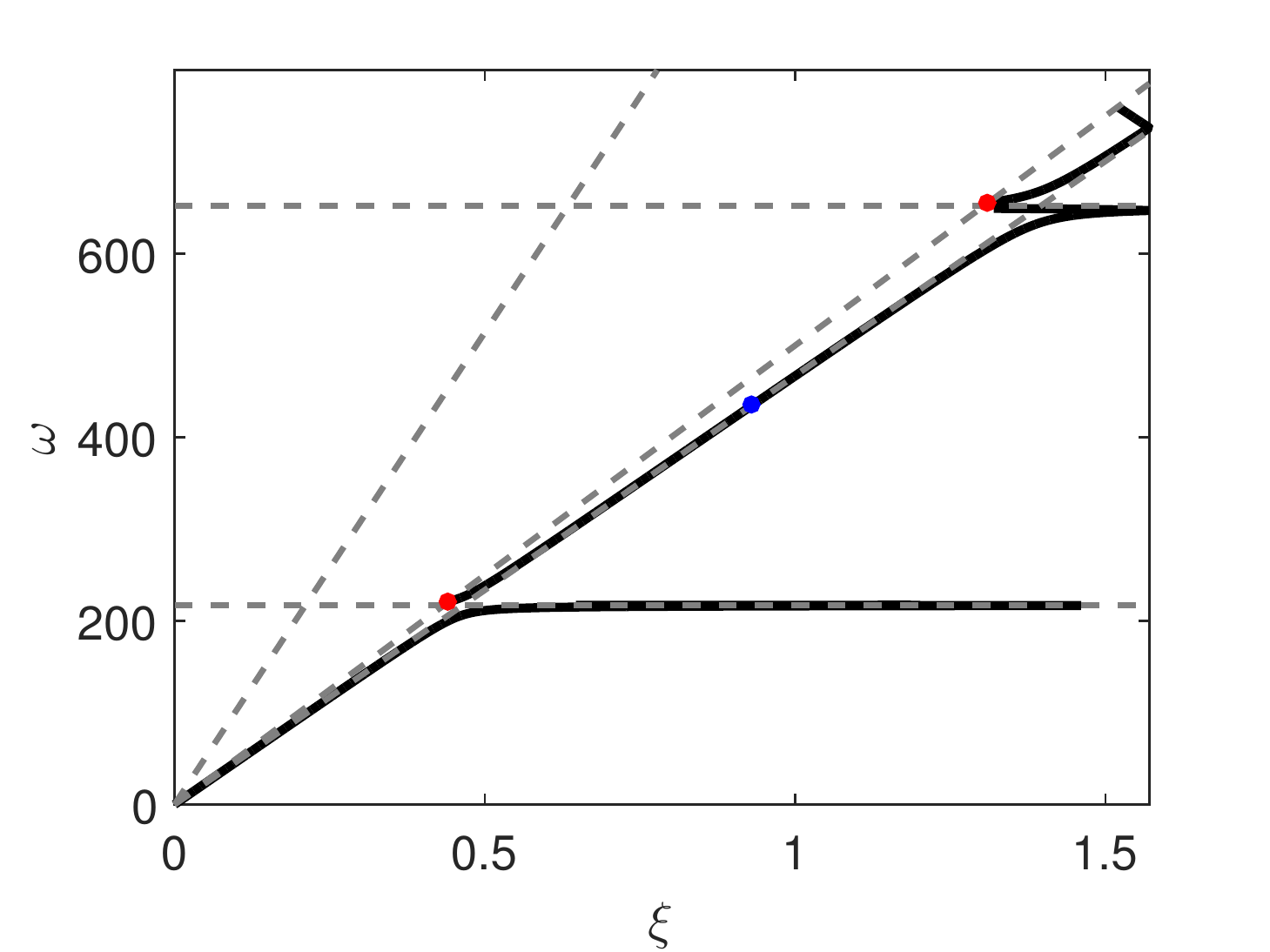}
\caption{\label{fig:half-plane-disp}
The dispersion curves for the array of resonators resting on the boundary of an elastic half-plane.
The solid black lines show the solutions of the dispersion equation,
the horizontal dashed lines indicate the resonances of the resonators,
the dashed grey lines of increasing slopes correspond to the
Rayleigh-wave, shear, and compressional sound-lines respectively.
The red dots indicate the intersection of the dispersion curves with the shear-wave sound-line whilst the blue dot corresponds to the intersection of the curves with the Rayleigh-wave sound-line.}
\end{figure}

\subsection{Dispersive properties and critical points}

\begin{table}[htb]
\centering
\begin{tabular}{c@{\qquad}c@{\qquad}c@{\qquad}}
\toprule
\multicolumn{2}{c}{Parameter} 			&			  \\
Symbol		&		 Definition 		&		 Value \\
\midrule
$L$			& Lattice spacing			& $2\;\text{m}$ \\
$\rho$		& Half-plane density		& $13000\;\text{kg}\cdot\text{m}^{-3}$\\
$\mu$		& Half-plane shear modulus	& $325\;\text{MPa}$\\
$\lambda$	& Half-plane first Lam\`e
				parameter				& $702\;\text{MPa}$\\
$\ell$		& Resonator length			& $14\;\text{m}$ \\
$\rho_R$	& Resonator density			& $450\;\text{kg}\cdot\text{m}^{-3}$\\
$h$			& Resonator diameter		& $0.3\;\text{m}$\\
$\mu$		& Resonator shear modulus	& $668\;\text{MPa}$\\
$\lambda$	& Resonator first Lam\`e
				parameter				& $780\;\text{MPa}$\\
\bottomrule
\end{tabular}
\caption{\label{tab:elastic-parameters}
The geometrical and numerical parameters used to produce the dispersion curves for the half-plane system shown in figure~\ref{fig:half-plane-disp}.}
\end{table}

Figure~\ref{fig:half-plane-disp} shows the dispersion curves for a typical configuration; the material parameters are detailed in table~\ref{tab:elastic-parameters}.
The dispersion curves, which are solutions of~\eqref{eq:reduced-dispersion}, are shown as solid black lines.
These curves correspond to combinations of frequency, $\omega$, and wavenumber, $\xi$, for which a surface waves exists; that is, waves that propagate over the surface of the half-space and decay exponentially into the bulk.
These waves arise as a result of the periodicity of the array present on the boundary of the half-space and are distinct from the usual Rayleigh waves that exist on free surfaces;
it is natural, therefore, to refer to such waves as Rayleigh-Bloch waves~\cite[]{porter99a,colquitt2015rayleigh}.

In addition to the dispersion curves, we also indicate the compressional resonances of the resonators, shown as the horizontal dashed lines in figure~\ref{fig:half-plane-disp}.
The remaining dashed lines of increasing slope are the curves corresponding to Rayleigh waves, bulk shear waves, and bulk compressional waves in the homogeneous elastic half-space respectively;
we call these dispersion lines the Rayleigh/shear/compressional wave sound-line by analogy with the terminology used in electromagnetism (see, for example,~\cite[]{joannopoulos2011photonic}).

Previous investigations of similar systems were based on purely numerical simulations~\cite[]{colombi16a,colombi15b}; with the dispersion equation~\eqref{eq:reduced-dispersion} in hand, we can now examine several interesting features of the dispersion curves offering additional physical insight.
As noted in the earlier papers~\cite[]{colombi14a,williams15a} on
elastic plates and from the numerical simulations of \cite{colombi15b}
for a half-space, the onset of band gaps coincide with the longitudinal resonances of the resonators.
Here we observe that the upper boundary of the band gaps coincide with the intersection of the dispersion curves with the shear sound-line.
These points are indicated by the red circles in figure~\ref{fig:half-plane-disp} and correspond to the case when $\xi=1$ ($\omega/k = v_s$), in which case, equation~\eqref{eq:reduced-dispersion} reduces to $L^2\sqrt{\mu\rho} + \sqrt{1 - r^2}S\sqrt{E\Rho}\tan(\ell\omega\alpha_R) = 0$ which has the solutions
\begin{equation}
\omega^{(s)}_n = \frac{1}{\ell\alpha_R}\left[n\pi - \arctan\left(\frac{L^2}{S\sqrt{1-r^2}}\sqrt{\frac{\mu\rho}{E\Rho}}\right)\right],\quad \text{for}\quad n\in\mathbb{N}.
\end{equation}
The longitudinal resonances of the resonators lie at $\omega^{(r)}_n = (2n-1)\pi/(2\ell\alpha_R)$.
Thus, in this regime where the pass bands are bounded from below by the resonances of the resonators and from above by the intersections of the dispersion curves with the shear sound-line, the band gaps have a constant width defined by
\begin{equation}
\delta\omega = \frac{1}{\ell\alpha_R}\left[\frac{\pi}{2} - \arctan\left(\frac{L^2}{S\sqrt{1-r^2}}\sqrt{\frac{\mu\rho}{E\Rho}}\right)\right].
\label{eq:elastic-band-width}
\end{equation}
Since $0 < r^2 < 3/4$, the argument of $\arctan$ in equation~\eqref{eq:elastic-band-width} is real and positive; hence $\delta\omega = \omega^{(s)}_n - \omega^{(r)}_n > 0$ and therefore band gaps always exist for finite parameters.

Equation~\eqref{eq:elastic-band-width} provides a convenient formula
for tuning the width of the band gap: 
In order to maximise the band gap, the factor $\ell\alpha_R$ and the argument of $\arctan$ in~\eqref{eq:elastic-band-width} should be made as small as possible simultaneously.
The converse will minimise the width of the stop band.
In rough terms, short resonators with high compressional wave-speeds will maximise the stop band width.
However, as the material parameters of the resonators also appear in the argument of $\arctan$, some care is required when tuning the system.

\subsection{Application to filters, metawedges, and effective band-gap widths}

Thus far we have studied infinite periodic structures.
However, metamaterials, metasurfaces, and phononic structures often find application as filters, lenses, and polarisers.
As has been shown in the electromagnetic~\cite[]{maradudin11a}, elastodynamic~\cite[]{colquitt2011dispersion}, and thin plate~\cite[]{sebbah_plate} literature, the dispersive
properties of an infinite system can be used to design finite
structures that posses interesting properties such as negative
refraction, flat lenses, filtering, and cloaking.
To illustrate the potential of Rayleigh waves in this regard, we consider their interaction with a finite array of resonators.

To this end, and given that we are concerned with the control of surface waves on elastic bodies, the intersection of the dispersion curves with the Rayleigh wave sound-line for an homogeneous elastic half-space without resonators (indicated by the blue dot on figure~\ref{fig:half-plane-disp}) is of interest.
For combinations of wavenumber and frequency corresponding to this point of intersection, a Rayleigh wave incident on the array of resonators from the elastic half-space can generate a Rayleigh-Bloch wave, that is a surface wave whose behaviour is affected by the periodicity, in the array.
Conversely, at this point, a Rayleigh-Bloch wave leaving the array can couple into a pure Rayleigh wave in the ambient half-space.
With the exception of embedded surface waves, which we do not consider here, the homogeneous half-space does not support localised Rayleigh waves corresponding to points above the Rayleigh wave sound-line;
in practical terms, this means that the region between the lower red dot and the blue dot on figure~\ref{fig:half-plane-disp} is an ``effective stop band'' for Rayleigh waves incident on the array from a homogeneous half-space.

\begin{figure}
\centering
\includegraphics[width=\linewidth,
trim={50 260 65 45},clip]{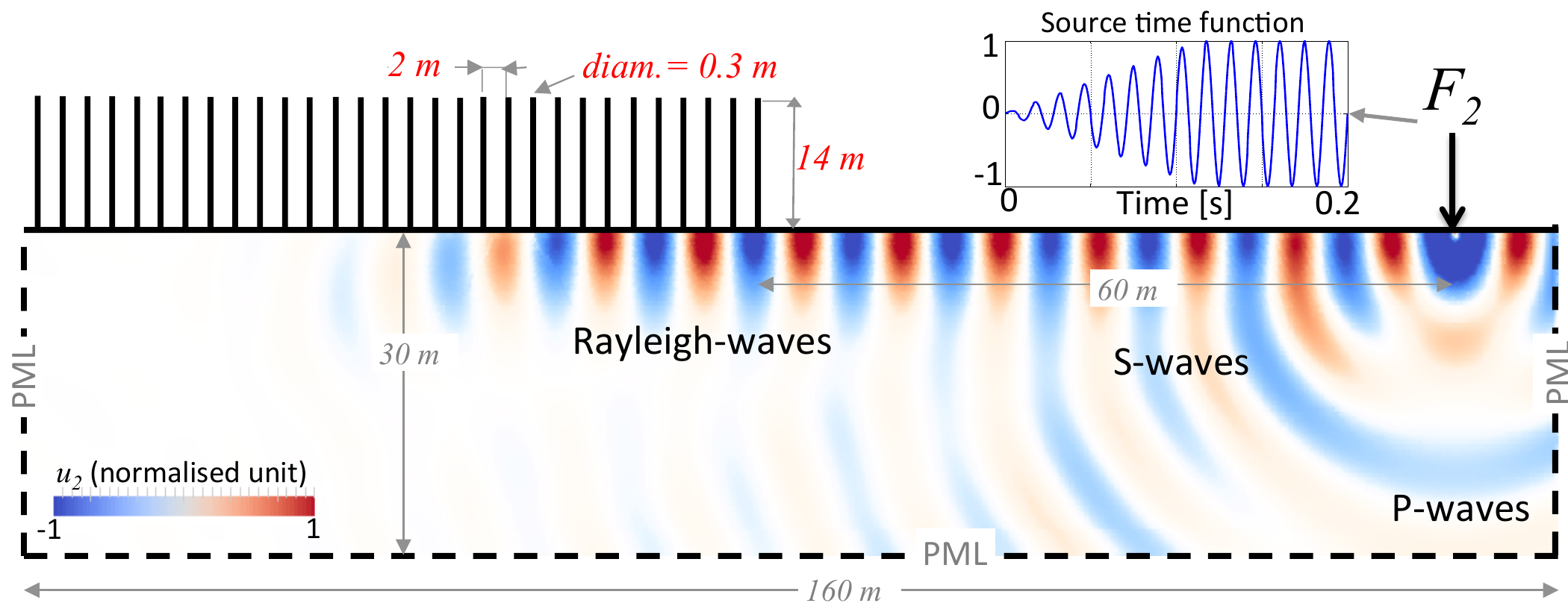}
\caption{\label{fig:half-line-fig0}
The 2D computational domain for the half-space also showing, in
a red-blue colorscale, the vertical displacement $u_{2}$ (proportions
are not to scale). The inset depicts the source time function generating the monochromatic Rayleigh wave.}
\end{figure}

The point of intersection between the dispersion curves and the Rayleigh sound-line can be obtained from~\eqref{eq:reduced-dispersion} by setting $\xi = v_r/v_s$, where $v_r$ is the Rayleigh wave speed (see~\cite[]{graff75a}, among others).
The frequency of intersection is then
\begin{equation}
\omega_n^{(R)} = \frac{1}{\ell\alpha_R}\left\{ n\pi - \arctan\left[\frac{1}{S}\sqrt{\frac{\mu\rho}{E\Rho}}\left( \frac{\left[2\gamma^2-1\right]^2}{\sqrt{\gamma^2 - r^2}} - 4\gamma^2\sqrt{\gamma^2 - 1}\right)\right]\right\},
\label{eq:rayleigh-intersection}
\end{equation}
where $\gamma = v_s/v_r$.
The ``effective width'' of the first band gap is then
\begin{equation}
\delta\omega^{(\text{eff})} = \frac{1}{\ell\alpha_R}\left\{ \frac{\pi}{2} - \arctan\left[\frac{1}{S}\sqrt{\frac{\mu\rho}{E\Rho}}\left( \frac{\left[2\gamma^2-1\right]^2}{\sqrt{\gamma^2 - r^2}} - 4\gamma^2\sqrt{\gamma^2 - 1}\right)\right]\right\}.
\end{equation}

\begin{figure}
\centering
\includegraphics[width=\linewidth,
trim={0 125 0 100},clip]{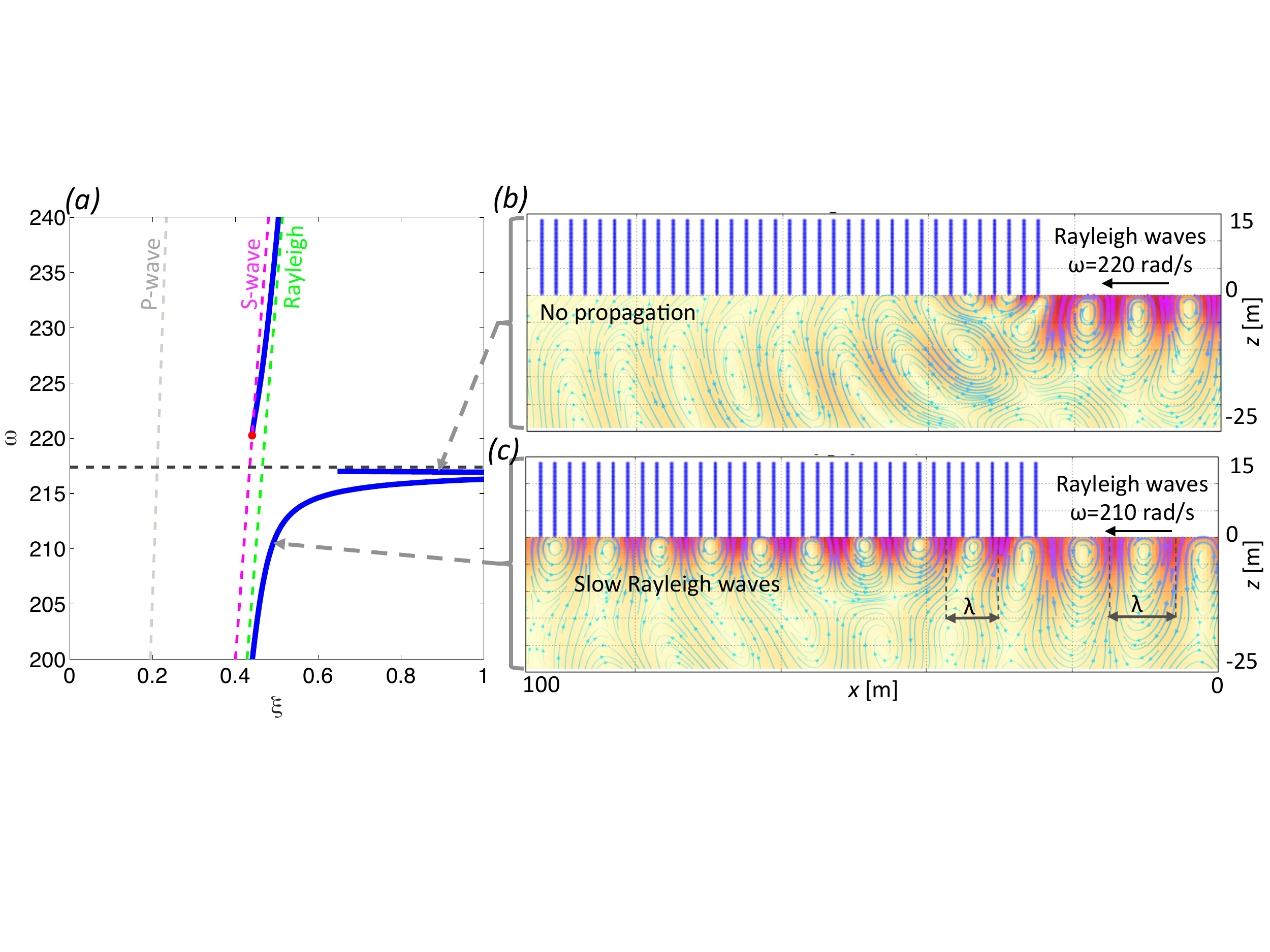}
\caption{\label{fig:half-line-fig1}
(a) The plot shows a magnified section of
figure~\ref{fig:half-plane-disp}, highlighting the first two branches
and the first resonance. P, S and Rayleigh wave dispersion curves for
the free half-space are also annotated in different colours.
(b,c)  Wavefield inside the half-space computed as described in figure~\ref{fig:half-line-fig0}.
(b) illustrates the stop band behaviour of the array for surface waves in the vicinity of a resonance, whilst (c) shows the usual pass band behaviour associated with the lower branch, $\lambda$ represents the wavelength. Elastic streamlines are superimposed to the wavefield magnitude colorcode. The field in the resonators is not shown.
An animated version of this figure can be found in the supplementary material{\protect\footnotemark[3]}.}
\end{figure}

\begin{figure}
\centering
\includegraphics[width=\linewidth,
trim={0 125 0 100},clip]{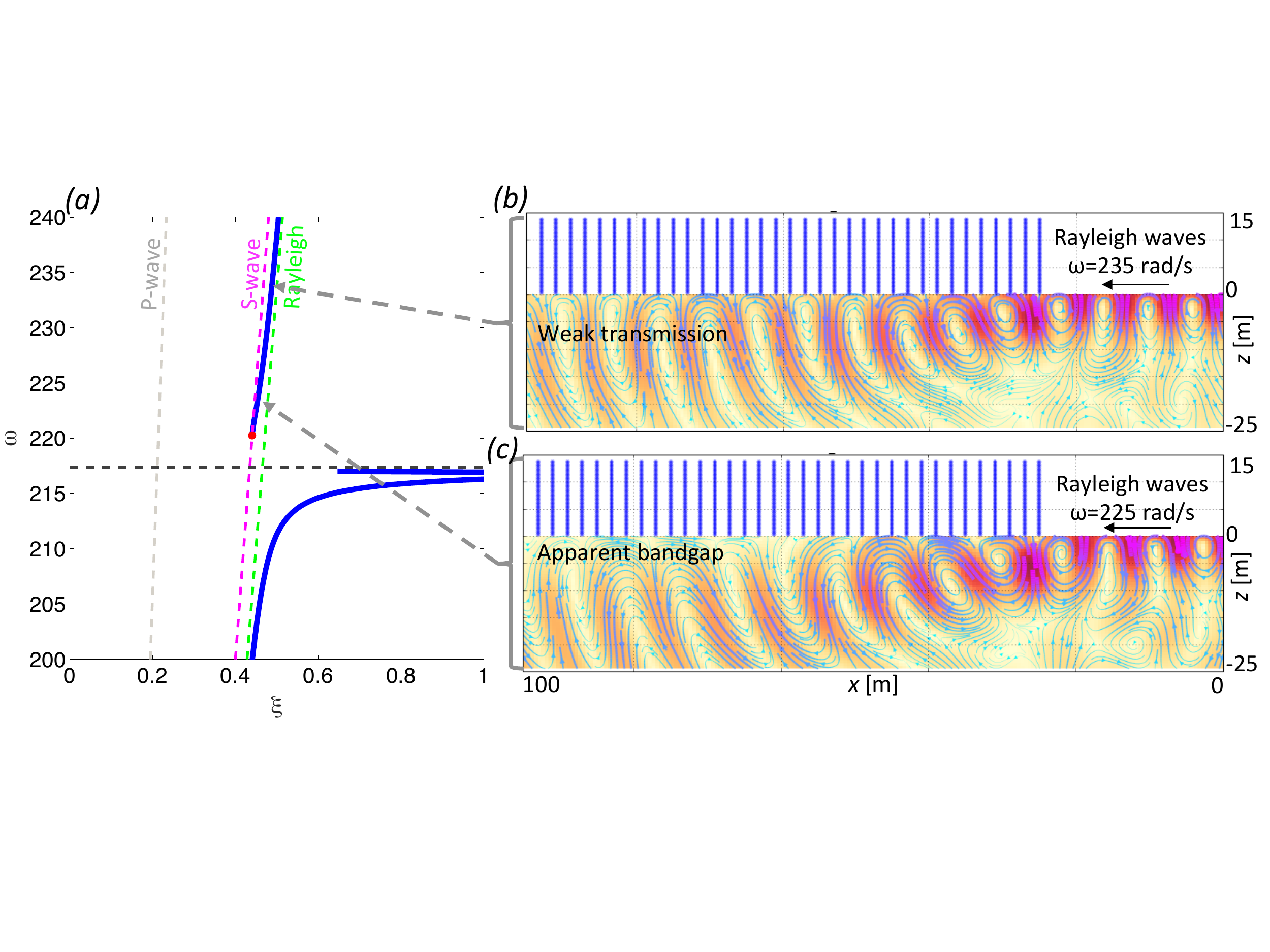}
\caption{\label{fig:half-line-fig2}
Same as figure~\ref{fig:half-line-fig1} but for the upper branch of the dispersion (blue) line.
(b) illustrates the weak transmission of Rayleigh waves for
frequency-wavenumber combinations just above the Rayleigh sound-line. (c) demonstrates the behaviour close to the intersection of the dispersion curves with the Rayleigh sound-line and responsible for the apparent bandgap.
An animated version of this figure can be found in the supplementary material{\protect\footnotemark[3]}}
\end{figure}

We now illustrate the dispersive properties of the resonant array atop an elastic half-space using the results from time domain spectral element (SEM) simulations.
We consider a linearly elastic, isotropic, homogeneous medium with a linear array of resonators, as depicted in figure~\ref{fig:half-line-fig0}; the material and goemetrical properties of the system are specified in table~\ref{tab:elastic-parameters}.
The simulations are performed with SPECFEM2D\footnote{\href{https://geodynamics.org/cig/software/specfem2d/}{https://geodynamics.org/cig/software/specfem2d/}} \cite[]{Komatitsch01041998}, a well known parallel code widely used by the seismological community.
The mesh is constructed using quadrilateral elements and the commercial software CUBIT\footnote{\href{https://cubit.sandia.gov/}{https://cubit.sandia.gov/}}.
Perfectly matched layers~\cite[]{Komatitsch_CPML} (PML) are
applied on the bottom and vertical boundaries of the computational domain containing the half-space; the remaining boundaries  (top surface and resonators) are traction free.
The accuracy of this method has been thoroughly tested in a previous study~\cite[]{colombi15b} against experimental measurements.

Thus far we have considered an infinite array of resonators, however for the numerical simulations
(Figs.~\ref{fig:half-line-fig1} and~\ref{fig:half-line-fig2}), a large, finite,
number is more convenient.
We take an array of 30 resonators and place a point source 60m away on the surface of the half-space, as shown in figure \ref{fig:half-line-fig0}.
The source takes the form of a sinusoidal vertical force of angular frequency $\omega$.
To avoid spurious effects at the onset of time, the amplitude of the sinusoidal force is initially modulated by a ramp function that progressively increases in amplitude over the first eight periods (see the inset of figure~\ref{fig:half-line-fig0}),
\begin{equation}
F(t)=
\begin{cases}
0 												& \text{for } t < 0, \\
\dfrac{\omega t}{16\pi}\sin(\omega t) 	& \text{for } 0\leq t < \dfrac{16\pi}{\omega}, \\
\sin(\omega t) 								& \text{for } t\geq \dfrac{16\pi}{\omega}.
\end{cases}
\end{equation}
The distance between the source and the resonant array is such that only Rayleigh waves scatter on the resonators (e.g. Figs.~\ref{fig:half-line-fig1} and~\ref{fig:half-line-fig2}); the remaining classes of waves are absorbed by the PML.
Once the ramp function has reached its final value, the simulations continue in a stationary state where only a monochromatic signal with constant amplitude propagates.
The snapshots in Figs.~\ref{fig:half-line-fig1} and \ref{fig:half-line-fig2} have been taken when the simulation is in this regime.
Figures~\ref{fig:half-line-fig1} and~\ref{fig:half-line-fig2} demonstrate the metamaterial properties of the resonant array in various regimes.
Here we have chosen not to present the deformation of the resonators in order to highlight the wavefield in the half-space; the full videos associated with these figures are provided in the supplementary materials\footnote{\href{http://dx.doi.org/10.6084/m9.figshare.3515000}{doi: 10.6084/m9.figshare.3515000}}.

Figure~\ref{fig:half-line-fig1}c shows the transmission of a surface wave through the array at a frequency corresponding to the lower branch of the dispersion curves, figure~\ref{fig:half-line-fig1}a.
We also note the decrease in wavelength inside the array corresponding
to the slower effective wavespeed inside the metasurface as compared with the homogeneous half-space.
The shorter wavelength is accompanied by an increase in amplitude of the waves inside the array as well as an increased rate of decay into the bulk.
In contrast, figure~\ref{fig:half-line-fig1}b shows the case when the frequency of the incident wave almost coincides with a resonance of the resonators leading to virtually no transmission.

Moving on to figure~\ref{fig:half-line-fig2}, the two panels (b) and (c)
correspond to the second branch of the dispersion curves, as indicated.
It is important to note that this upper branch transitions
from the Rayleigh sound-line to the shear sound-line, that is, the Rayleigh-Bloch wave in the resonant array preserves the Rayleigh-like surface wave
properties at higher frequencies but will evolve into a
shear-like wave as the frequency decreases. 
The hybrid nature of this branch, sitting between the shear and Rayleigh
sound-lines, is responsible for the mode conversion of Rayleigh waves into
shear waves and used in the design of so-called resonant \emph{metawedge} systems \cite[]{colombi16a}.

The point of intersection (marked in red) between the hybrid branch
and the shear sound-line is of particular interest.
At this frequency, waves can propagate past the metasurface but they must only do so as shear waves (note the polarization) as is illustrated in figure~\ref{fig:half-line-fig2}(c) for a frequency very close to this point.
In this regime the transition between Rayleigh waves propagating in the free half-space (to the right hand side of the array) and the hybrid
mode in the array happens abruptly resulting in an ``apparent bandgap"
for surface waves.
Therefore the wavefield beneath the resonators, represented in figure~\ref{fig:half-line-fig2}(c), is characterised by shear waves that propagate away from the surface and into the bulk leaving the left hand side of the array almost untouched.
As the frequency increases (figure~\ref{fig:half-line-fig2}(b)) the wavefield reveals more clearly its hybrid nature.
The waves propagating under the resonators are grazing the surface but they are not yet exponentially decaying as Rayleigh waves.
In this regime waves can propagate in the resonator but the transmission is still weak.
The hybrid mode progressively turns into a Rayleigh wave as the frequency increases toward the crossing point (blue dot in figure~\ref{fig:half-plane-disp}). 

\section{Concluding remarks}
\label{sec:conclude}

Metamaterials barely existed fifteen years ago, but now form a major
research area: although initially developed for optics, the field of metamaterial research has expanded rapidly and now includes the development of novel materials for applications in acoustics and elasticity.
More recently, the applicability of metamaterials to seismology has sparked the interest of geophysicists in the development of novel methods to control surface waves.
Given the interest in this emerging area, there is a need to study the properties of the solutions to fundamental canonical problems.

In this paper we have provided the theoretical and analytical
framework necessary for the rigorous study  of these seismic
metasurfaces that adjust the surface wave behaviour.
The framework is introduced through the study of two canonical problems associated with the control of mechanical surface waves.
Initially, we study the propagation of surface Bloch waves through an array of resonators on a thin elastic plate.
In this case, the resonators are thin elastic rods supporting both compressional and flexural waves.
Explicit exact solutions are developed and used to examine the behaviour of the system.
In particular, we study the coupling between the symmetric and anti-symmetric modes in the substrate and the resonances of the rods.
Interestingly it is found that, in the frequency range of interest and for sufficiently rigid plates, the flexural resonances of the resonators couple very weakly into the substrate with the dominant effect coming from the compressional modes of resonators, which open up band gaps associated with the compressional resonances.
For more flexible plates, the flexural deformations of the resonators become important.

The plate system is used as the motivation for the far more challenging problem considered in section~\ref{sec:full_elastic} where we examine the propagation of surface Bloch waves through an array of resonators resting on a fully-elastic half-space.
Here we develop closed form expressions for the dispersion equation and wave-fields in the deep sub-wavelength regime of interest.
As for the plate problem, the compressional resonances of the resonators create band gaps in the dispersion curves and analytical expressions for their position and width are provided.
In particular, it is shown that band gaps will always exist regardless of the material or geometrical parameters of the half-space and resonators.
The notion of ``effective band gaps'' for surface waves is also introduced and discussed. 

In section~\ref{sec:full_elastic}, we also examine the scattering problem associated with a half-line of resonators and use the dispersive properties of the array to examine, the filtering effects of the resonant array.

With the formal framework of seismic metasurfaces now established, the time is ripe to exploit these results and explore the possibilities of extending the existing results for electromagnetism and acoustics to seismology.
Indeed, the theoretical framework developed in the present paper is
already being used in the design and development of the so-called
\emph{meta-wedge} that is capable of mode-converting destructive
seismic surface waves into mainly harmless bulk shear waves~\cite[]{colombi16a}.

\section*{Acknowledgements}
The authors thank the EPSRC (UK) for their support through the Programme Grant EP/L024926/1.
P.R. acknowledges the financial support of ANR under project LabEx OSUG\at2020.

\appendix 
\section{The dispersion equation for the plate with flexural resonators}
\label{app:dispersion-plate}

The dispersion equation~\eqref{eq:plate-disp2} for Bloch-waves propagating through the plate can be cast as a cubic polynomial in $\cos\xi$ with the coefficients depending on angular frequency $\omega$ as well as the material and geometrical parameters.
These coefficients are cumbersome and are therefore stated here,
rather than disturb the flow of the main text.
With reference to equation~\eqref{eq:plate-disp2}, the polynomial coefficients are as follows
\begin{equation}
A_3(\omega) = - 32\beta^5\alpha\omega^{7/2},
\end{equation}
\begin{multline}
A_2(\omega) = 
8\beta^2\omega^2\left\{ 4\beta^3\alpha\omega^{3/2}\left[ \cos\beta\sqrt{\omega} + \cos\alpha\omega \right]  + 4\beta^3 \alpha\omega^{3/2}\cosh\beta\sqrt{\omega}
\right. \\
\left.
- \alpha\left[ V + \beta^2 M_\theta\omega \right] \sin\beta\sqrt{\omega} + 2 \beta^3 F_u\sqrt{\omega}\sin\alpha\omega
\right.
\\
\left.
+ \alpha\left[ V - \beta^2 M_\theta\omega\right]\sinh\beta\sqrt{\omega} \right\},
\end{multline}
\begin{multline}
A_1(\omega) = -4\beta\omega\left\{
\alpha M_\theta v^{(\omega)}\sqrt{\omega} + 4\beta^4\omega^{3/2}\cos\beta\sqrt{\omega}\left[2\alpha\omega\cos\alpha\omega + F_u\sin\alpha\omega \right]
\right.
\\
+ \sqrt{\omega}\cosh\beta\sqrt{\omega}\left[ \alpha\left(8\beta^4\omega^2 - M_\theta V\right)\cos\beta\sqrt{\omega} + 8\beta^4\alpha\omega^2\cos\alpha\omega
\right.
\\
\left.
- 2\beta\alpha\sqrt{\omega}\left(V + \beta^2\omega M_\theta\right)\sin\beta\sqrt{\omega} +4\beta^3 F_u\omega\sin\alpha\omega \right]
\\
- \beta\sin\beta\sqrt{\omega}\left[ \left( F_uV + \beta^2\omega\left\{ F_uM_\theta - F_\theta M_u \right\}\right)\sin\alpha\omega
\right.
\\
\left.
+ 2\alpha\omega\left( V + \beta^2\omega M_\theta \right)\cos\alpha\omega
\right]
+
\beta\left[ 2\alpha\omega\left( V - \beta^2\omega M_\theta \right)\left(\cos\beta\sqrt{\omega} + \cos\alpha\omega \right)
\right.
\\
\left.
+
F_u V + \beta^2\omega\left\{ F_\theta M_u - F_uM_\theta\right\}\sin\alpha\omega \right]\sinh\beta\sqrt{\omega},
\end{multline}
\begin{multline}
A_0(\omega) = 
\beta\sqrt{\omega}\cos\beta\sqrt{\omega}\cosh\beta\sqrt{\omega}\left\{
2\alpha\omega\left[ 16 \beta^4 \omega^2 - M_\theta V \right]\cos\alpha\omega + \left[ 16\beta^4\omega^2 F_u 
\right.\right.
\\
\left.\left.
+ V\left(F_\theta M_u - F_uM_\theta\right)\right]\sin\alpha\omega
+ 2V\sec\beta\sqrt{\omega}\sech\beta\sqrt{\omega}\left[ 2\alpha M_\theta\omega\cos\alpha\omega
\right.\right.
\\
\left.
\left.
+ \left(F_uM_\theta - F_\theta M_u\right)\sin\alpha\omega\right]
+ V\sech^2\beta\sqrt{\omega}\left[ \sin\alpha\omega\left(M_uF_\theta
\right.\right.\right.
\\
\left.\left.\left.
- F_uM_\theta\right) - 2\alpha\omega M_\theta\cos\alpha\omega \right]
- 4\beta\sqrt{\omega}\left[ 2\alpha\omega\left( V + \beta^2\omega M_\theta \right)\cos\alpha\omega 
\right.\right.
\\
\left.\left.
+ \left( F_u V + \beta^2\omega \left\{ F_u M_\theta - F_\theta M_u \right\}\right)\sin\alpha\omega\right]\tan\beta\sqrt{\omega}
\right.
\\
\left.
+ 4 \beta \sqrt{\omega}\left[ 2\alpha\omega\left( V - \beta^2\omega M_\theta\right)\cos\alpha\omega + \left( F_uV + \beta^2\omega\left\{ F_\theta M_u
\right.\right.\right.\right.
\\
\left.\left.\left.\left.
- F_u M_\theta\right\}\right)\sin\alpha\omega \right]\tanh\beta\sqrt{\omega}
+ V\left[ \sin\alpha\omega\left( F_\theta M_u - F_uM_\theta \right)\sin\alpha\omega
\right.\right.
\\
\left.\left.
- 2\alpha M_\theta\omega\cos\alpha\omega \right]\tanh^2\beta\sqrt{\omega}\right\}.
\end{multline}

\section{The dispersion equation for the plate with rod-like resonators}
\label{app:dispersion-plate-long}

The dispersion equation~\eqref{eq:plate_disp_long} for flexural waves propagating through a plate with rod-like resonators can be expressed as quadratic polynomial in $\cos\xi$ with coefficients depending on the spectral properties of the plate and resonators.
These coefficients are
\begin{equation}
B_2(\omega) = -4\beta^4\omega^{3/2},
\end{equation}
\begin{equation}
B_1(\omega) = 4\beta^3\omega^{3/2}\left[\cos\beta\sqrt{\omega} + \cosh\beta\sqrt{\omega}\right] + V\left[\sinh\beta\sqrt{\omega} - \sin\beta\sqrt{\omega}\right],
\end{equation}
\begin{equation}
B_0(\omega) = \left[V\left(\tanh\beta\sqrt{w} - \tan\beta\sqrt{\omega}\right) - 4\beta^3\omega^{3/2}\right]\cos\beta\sqrt{\omega}\cosh\beta\sqrt{\omega}.
\end{equation}

\end{document}